%% file: main.tex
\documentclass[sigconf,10pt,nonacm]{acmart}
\input{preamble}

\begin{document}

\title{\DesignName: Disk-aware KV Cache Offloading for Long-Context On-device Inference }

\author{Huawei Zhang}
\email{schz@leeds.ac.uk}
\affiliation{%
  \institution{University of Leeds}
  \city{Leeds}
  \country{UK}
}

\author{Chunwei Xia}
\email{C.Xia@leeds.ac.uk}
\affiliation{%
  \institution{University of Leeds}
  \city{Leeds}
  \country{UK}
}

\author{Zheng Wang}
\email{Z.Wang5@leeds.ac.uk}
\affiliation{%
  \institution{University of Leeds}
  \city{Leeds}
  \country{UK}
}

\renewcommand{\shortauthors}{}

\input{sections/0.abstract}

\maketitle

\input{sections/1.introduction}

\input{sections/2.background}

\input{sections/3.motivation}

\input{sections/4.design}

\input{sections/5.0.setup}

\input{sections/5.experiment}

\input{sections/6.analysis}

\input{sections/7.related_work}

\input{sections/8.conclusion}


\bibliographystyle{ACM-Reference-Format}
\bibliography{ref}
\balance

\input{sections/9.appendix}

\end{document}

%% file: preamble.tex
\usepackage{makecell}
\usepackage{graphicx}
\usepackage{subcaption}
\usepackage{xspace}
\usepackage{multirow}
\usepackage{enumitem}
\usepackage{float}
\usepackage[export]{adjustbox}
\usepackage[ruled,vlined]{algorithm2e}
\usepackage{lineno}
\usepackage{enumitem}
\usepackage{listings}
\usepackage{siunitx}
\usepackage[table]{xcolor}
\usepackage{balance}
\usepackage{siunitx}

\newcommand{\DesignName}{\textsc{KVSwap}\xspace}
\newcommand{\SystemName}{\DesignName}

\newcommand{\cparagraph}[1]{\vspace{0.5mm}\noindent\textbf{#1}}

\definecolor{keywordcolor}{RGB}{0,0,255}  
\definecolor{stringcolor}{RGB}{163,21,21}
\definecolor{commentcolor}{RGB}{0,128,0}    
\definecolor{numbercolor}{RGB}{163,21,21}  
\definecolor{bgcolor}{RGB}{255,255,255}

\lstdefinestyle{xcodestyle}{
    language=Python,
    basicstyle=\ttfamily\small,       
    keywordstyle=\color{keywordcolor}\bfseries,
    commentstyle=\color{commentcolor}\itshape,
    stringstyle=\color{stringcolor},
    numbers=none,                       
    frame=none,                         
    breaklines=false,                    
    tabsize=4,                          
    showstringspaces=false,
    escapeinside=||,
    mathescape=true,
    columns=fullflexible,
    backgroundcolor=\color{bgcolor},   
    literate=*                     
      {0}{{{\color{numbercolor}0}}}1
      {1}{{{\color{numbercolor}1}}}1
      {2}{{{\color{numbercolor}2}}}1
      {3}{{{\color{numbercolor}3}}}1
      {4}{{{\color{numbercolor}4}}}1
      {5}{{{\color{numbercolor}5}}}1
      {6}{{{\color{numbercolor}6}}}1
      {7}{{{\color{numbercolor}7}}}1
      {8}{{{\color{numbercolor}8}}}1
      {9}{{{\color{numbercolor}9}}}1
}

\definecolor{lightblue}{RGB}{220,230,255}
\definecolor{lightgreen}{RGB}{220,255,220}
\definecolor{lightpurple}{RGB}{237,224,255}

\newcolumntype{Z}{S[table-format=+3.2, explicit-sign]}

%% file: sections/0.abstract.tex
\begin{abstract}

Language models (LMs) underpin emerging mobile and embedded AI applications like meeting and video summarization and document analysis, which often require processing multiple long-context inputs. Running an LM locally on-device improves privacy, enables offline use, and reduces cost, but long-context inference quickly hits a \emph{memory capacity wall} as the key-value (KV) cache grows linearly with context length and batch size. Existing KV-cache offloading schemes are designed to transfer cache data from GPU memory to CPU memory; however, they are not suitable for embedded and mobile systems, where the CPU and GPU (or NPU) typically share a unified memory and the non-volatile secondary storage (disk) offers limited I/O bandwidth. We present \DesignName, a software framework tailored for local devices that achieves high memory efficiency while effectively leveraging disk storage. 
\SystemName stores the full cache on disk, uses highly compact in-memory metadata to predict which entries to preload, overlaps computation with hardware-aware disk access, and orchestrates read patterns to match storage device characteristics. 
Our evaluation shows that across representative LMs and storage types, \DesignName delivers higher throughput under tight memory budgets while maintaining generation quality over existing KV cache offloading schemes.

\end{abstract}

%% file: sections/1.introduction.tex
\section{Introduction}
Language models (LMs) have demonstrated strong capabilities in long-form document analysis and multimedia understanding~\cite{zhang2025systematic,laskar2023building,yin2024survey}. These underpin emerging mobile and embedded applications such as smart note-taking for meetings~\cite{otterai2025,firefliesai2025}, document analysis~\cite{adobeacrobatai2025,gmailgemini2025}, voice assistants for transcription and summarization~\cite{appleintelligence2025,googlerecorder2025}, and video search with natural language queries~\cite{appleintelligence2025,google_lens_video_search2025}. Increasingly, users expect these capabilities to run directly on their devices.

On-device deployment offers clear benefits: enhanced privacy by keeping data on-device, offline functionality under poor connectivity, and avoiding LM API costs. With the growing availability of powerful GPUs and NPUs on modern systems on chips~\cite{qualcomm_snapdragon8elite,orangepi5pro,mediatek_dimensity9400plus}, local deployment is becoming feasible. 

However, running LMs efficiently on resource-constrained devices like mobile and embedded AI systems remains challenging due to limited memory capacity and I/O bandwidth. While quantization~\cite{frantar2022gptq,lin2024awq} and parameter offloading~\cite{alizadeh2023llm,xue2024powerinfer} reduce model footprint, a critical bottleneck lies in managing the key-value (KV) cache. Unlike fixed model weights, the KV cache grows linearly with sequence length and batch size, often surpassing model size (as shown in Fig.~\ref{fig:mem_dist}). On a 4B-parameter LM, batched inputs over 16K tokens already push the KV cache beyond device memory, causing latency spikes or out-of-memory failures. As LMs are increasingly applied to long-context tasks, practical on-device inference requires efficient KV cache management.

Prior work reduces KV cache overhead by offloading from GPU to CPU memory~\cite{lee2024infinigen,sun2024shadowkv}, alleviating GPU pressure in server settings with abundant CPU RAM. These techniques, however, are ill-suite for mobile and embedded systems, where GPUs/NPUs typically share only 8-32 GB of RAM with the CPU. A natural alternative is to offload the KV cache to non-volatile storage\footnote{We use \textit{disk} for non-volatile storage and \textit{memory} for main memory.}, such as NVMe-, UFS-, or eMMC-based devices. However, without careful optimization, this approach severely degrades performance: mobile memory bandwidth is roughly 100 GB/s, while disk bandwidth is often just ~1 GB/s - two to three orders of magnitude lower than their server counterparts.

\emph{Can we offload the KV cache to disk for long-context, on-device inference without sacrificing throughput or generation quality?} In answer, we propose \DesignName, a framework for extended-context LM inference on resource-constrained devices. Built on FlexGen~\cite{sheng2023flexgen}, \DesignName supports diverse LM architectures and storage types, with simple APIs to use.
\DesignName offloads KV cache entries to disk and partially reloads them during generation. 
It stores the complete KV cache on disk but maintains a highly compact in-memory K cache representation to predict which entries on the disk are critical for each layer. These entries are prefetched into a buffer ahead of computation, reducing disk stalls without compromising generation quality. The \SystemName runtime automatically overlaps disk transfers with compute and synchronizes memory-disk copies.

The design of \SystemName needs to overcome two major challenges introduced by on-device disk offloading. First, the ever-increasing KV-cache size introduced by long-context is a challenge for resource-constrained mobile and embedded devices. The limited memory capacity and strict per-application memory quotas result in a \emph{tight memory budget}, far beyond what existing KV-cache offloading methods, which were primarily designed for server-side deployments, can accommodate. 
To address this, \DesignName proposes a compression and reconstruction algorithm that enables aggressive memory compression while still achieving information recovery with adequate quality.

Second, embedded storage devices based on NVMe, UFS, and eMMC often exhibit \emph{read amplification}, where the controller reads larger physical units (e.g., entire NAND pages) than requested~\cite{haas2023modern,mohan2017analyzing}, resulting in extra data movement and unnecessary read operations. 
To mitigate this, \DesignName groups KV entries at appropriate granularities and optimizes access patterns. It also integrates a software cache (reuse buffer) to retain recently accessed entries across generations. This reduces redundant I/O, which is particularly important for low-bandwidth devices ($<200$~MB/s) and batched inference, where data movement dominates latency. 
Together, these specified designs and optimizations enable \DesignName for efficient long-context on-device inference with negligible loss of generation quality.

Recent work explores storing the full KV cache on disk with selective retrieval~\cite{liu2024cachegen,yao2025cacheblend,chen2025impress}. These methods, designed for cloud-serving scenarios, primarily optimize time-to-first-token (TTFT) to reduce inference start-up delay. However, they are ineffective on resource-constrained devices during iterative decoding, as they still incur high memory overhead and assume static rather than dynamically evolving KV caches.
\DesignName addresses this gap by \emph{targeting the decoding stage} of LM inference, efficiently managing dynamically changing KV caches to reduce memory footprint while preserving generation quality. It is the first disk-based KV cache offloading method that enables high-quality, long-context LM generation on resource-constrained platforms.

We evaluate \SystemName on text, reasoning and video LMs across model sizes, memory budgets, disk types, batch sizes, and context lengths. Compared with the strongest KV cache offloading baseline, \DesignName delivers up to 1.8$\times$ (NVMe) and 4.1$\times$ (eMMC) throughput improvements at 32K context length under the same tight memory budget, while also providing \textit{substantially higher} generation quality. Against the industry-grade vLLM~\cite{kwon2023efficient} - in an idealized setting where all available memory is dedicated to the KV cache - \DesignName achieves comparable or higher throughput while using 11.0$\times$ less KV cache memory. These results demonstrate that \DesignName offers high memory efficiency and throughput with preserved generation quality for on-device long-context LM inference.

This paper makes the following contributions. It presents: 
\begin{itemize}[topsep=3pt, leftmargin=*]
    \item the first disk-based KV cache offloading framework for efficient long-context, on-device LM decoding;
    \item a memory-efficient algorithm and system co-design that accurately identifies and accesses groups of critical KV entries, achieving substantial memory compression with preserved quality while mitigating disk read amplification;
    \item a runtime system for managing and caching KV entries, with optimized computation pipeline and reduced I/O transfers.
\end{itemize}

%% file: sections/2.background.tex
\section{Background and Motivation}

\subsection{Language Models\label{sec:lm}} 
Language models (LMs) process queries through a two-stage pipeline: \emph{prefilling} and \emph{decoding}.
During prefilling, the model reads the entire input prompt to generate the first output token.
During decoding, the model produces subsequent tokens one at a time, each conditioned on the previously generated tokens.
Prefilling happens only once, whereas decoding is repeated for every output token. For open-ended tasks like document summarization or dialogue, decoding can involve \emph{hundreds or even thousands} of steps, and this becomes even more common when using reasoning-focused models with chain-of-thought (CoT)~\cite{wei2022chain}.

Attention is central to modern LMs, where each token is represented by a
\emph{Query} (Q), \emph{Key} (K), and \emph{Value} (V) vector.
Scores are computed by comparing queries with keys, and the resulting weights
are applied to values to form contextualized representations.
\emph{Multi-head attention (MHA)}~\cite{vaswani2017attention} extends standard attention by computing multiple sets of QKV vectors in parallel, but increases memory use since each head
stores its own keys and values.
\emph{Grouped-query attention (GQA)}\cite{ainslie2023gqa}
addresses this by sharing keys and values across heads, reducing memory overhead
while maintaining accuracy. GQA is widely adopted in SOTA models such as LLaMA3
and Qwen3~\cite{dubey2024llama,yang2025qwen3}.

During generation, LMs use a \emph{key-value (KV) cache} to store keys and values
from past steps, enabling the reuse of computations and reducing inference latency~\cite{pope2023efficiently}.

%% file: sections/3.motivation.tex
\begin{figure}
  \centering
  \captionsetup{skip=0pt}
  \includegraphics[width=0.85\linewidth]{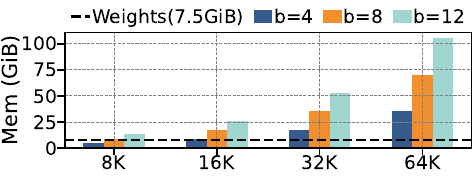}
  \caption{KV cache memory footprint across batch sizes (b) and sequence lengths (x-axis) of Qwen3-4B.}
  \label{fig:mem_dist}
   \vspace{-2mm}
\end{figure}

\subsection{On-device KV Cache Memory Wall}
The KV cache serves as a critical component of program state in on-device applications of LMs and is therefore usually required to be memory-resident. 
Fig.~\ref{fig:mem_dist} shows the KV cache memory footprint for the Qwen3-4B model (W16A16) under varying batch sizes and context lengths. The model weights alone occupy 7.5 GiB of RAM. At a context length of 16 K and batch size 4, the KV cache reaches 9 GiB, already exceeding the weight size. Because KV cache size scales linearly with both batch size and context length, it grows rapidly: at 32 K context and batch size 12, it reaches 54 GiB. 

On embedded and mobile systems with limited memory capacity (e.g., 8~GiB) and strict per-application quotas (to avoid disrupting other applications and OS services), retaining the full KV cache in memory becomes infeasible. This results in a \emph{KV cache memory capacity wall}, where the KV cache overtakes model weights as the dominant memory consumer, limiting the scalability of long-context on-device inference.

\subsection{KV Cache Disk Offloading}
\label{sub_src:disk_io_util}

\begin{figure}[t!]
  \centering
  \captionsetup{skip=0pt}
  \includegraphics[width=0.85\linewidth]{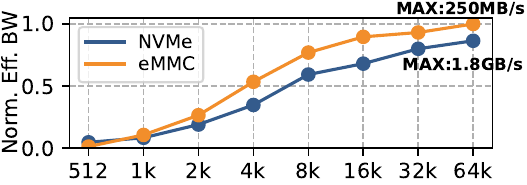}
  \caption{Normalized effective bandwidth (BW) with varying block sizes (bytes).}
  \label{fig:disk_bw_utils}
\end{figure}

Prior work shows that a small fraction of tokens substantially influence attention score computation, and that these influential tokens change dynamically with the context and query~\cite{zhang2023h2o,lee2024infinigen,sun2024shadowkv,tang2024quest}. This insight enables selective KV cache loading by predicting the most relevant tokens at each generation step. Existing techniques exploit this by offloading KV cache from GPU to CPU memory ~\cite{lee2024infinigen,sun2024shadowkv}, where both memory hierarchies are high-bandwidth and optimized for low-latency access. In contrast, our work targets embedded and mobile platforms, where RAM is scarce and subject to strict quotas. We propose offloading KV cache from memory to disk, targeting a different storage hierarchy with far more severe bandwidth and latency constraints.

To assess the feasibility of disk offloading, we profile the random read bandwidth of two representative types of disks - NVMe\footnote{UFS-based storage is not supported on our platform, but it exhibits I/O bandwidth and characteristics similar to NVMe.} and eMMC - under varying block sizes. 
We focus on random, rather than sequential access patterns because our setting requires \emph{selectively} retrieving critical KV entries, operations that inherently produce predominantly random accesses. From Fig.~\ref{fig:disk_bw_utils}, we observe:

\cparagraph{Diverse disk I/O bandwidths.}
 Disk types vary significantly in I/O bandwidth: NVMe can reach up to 1.8 GB/s, while eMMC can be limited to around 250 MB/s. 

\cparagraph{Poor I/O utilization for small KV entries.}  
All storage types show severe bandwidth under-utilization for small transfer sizes. At 512 bytes, the typical size of a KV entry\footnote{128 (head dimension) $\times$ 2 (key and value) $\times$ 2 Bytes (float16)  = 512 Bytes.}, the effective bandwidth drops below 6\% of peak bandwidth for both devices. 

We observe that bandwidth utilization improves substantially with larger block sizes. \DesignName addresses this by adopting a quality-aware, \emph{group-wise} strategy that packs multiple KV entries into a single transfer, thereby amortizing I/O overhead and reducing fragmentation while incurring negligible model accuracy degradation (Sec.~\ref{subsubsec:group_kv_pred}).

\subsection{Drawbacks of Prior Offloading Schemes}
\begin{figure}
  \centering
  \captionsetup{skip=0pt}
  \includegraphics[width=0.95\linewidth]{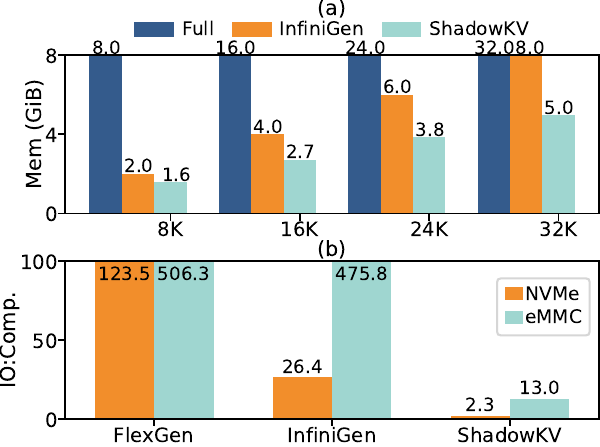}
    \caption{
        (a) KV cache management memory with batch=8 and varying context lengths (x-axis);
        (b) Decoding latency ratios of I/O to compute under 32K context length and batch=8.
    }
  \label{fig:mem_comparison}
\end{figure}

\cparagraph{High memory overhead.}
Recent KV cache offloading methods~\cite{sheng2023flexgen,lee2024infinigen,sun2024shadowkv} are designed for CPU-GPU transfer on servers and are ill-suited for disk-based on-device offloading. They do not explicitly consider extremely tight memory constraints (e.g., <500 MiB), a regime that closely matches the practical application memory budget in common on-device deployment scenarios.
Figure~\ref{fig:mem_comparison}a compares the KV cache management memory of InfiniGen~\cite{lee2024infinigen} and ShadowKV~\cite{sun2024shadowkv} against storing the full cache in memory for LLaMA3-8B. Their designs introduce a significant memory footprint, particularly for long contexts. For example, with a 16K context length and batch size 8, memory consumption already reaches 4~GiB and 2.7~GiB. 
Here, we attempted to reconfigure their parameters for tighter memory budgets and extend them to disk-based offloading, but this severely degrades generation quality (Sec.~\ref{subsec:gen_quality}). We attribute this to the inability of their memory compression algorithms to tolerate high compression ratios. A detailed discussion is provided in Sec.~\ref{sec:kcompressed} and Sec.~\ref{sec:gckvp}.

\cparagraph{Low I/O efficiency.}
The fine-grained head- or token-level operations create fragmented disk access patterns, causing excessive small disk reads and degrading I/O efficiency. 
As an example, Figure~\ref{fig:mem_comparison}b shows the decoding latency ratio of I/O to compute for FlexGen~\cite{sheng2023flexgen}, InfiniGen~\cite{lee2024infinigen}, and ShadowKV~\cite{sun2024shadowkv}. Note that FlexGen does not select KV entries; instead, it loads the full KV cache into memory. All these methods exhibit ratios far exceeding 1, with some cases reaching over 100. While ShadowKV achieves the lowest ratios, it still reaches 13.0 on eMMC and 2.3 on NVMe. These high ratios indicate a severely imbalanced I/O–compute pipeline caused by inefficient I/O access. End-to-end results in Sec.~\ref{sec:eval:throughput} further validate the resulting low system throughput for these methods.

\cparagraph{Lessons learnt.}
For on-device disk offloading, existing KV cache offloading methods suffer from \emph{high memory overhead}, which limits scalability to longer contexts or larger batch sizes.
In addition, their \emph{low I/O efficiency} significantly limits system decoding, resulting in low generation speed and energy waste. \DesignName is designed to avoid these pitfalls.

%% file: sections/4.design.tex
\begin{figure}[t]
\centering
\begin{subfigure}{0.95\linewidth}
\centering
\begin{lstlisting}[style=xcodestyle]
import KVSWAP
KVSWAP.Parameter_Tuning(
    model=model_obj, max_context_len=32768,
    max_batch_size=8, max_kv_mem=2200 # in MiB
).save("config.json")
\end{lstlisting}
\captionsetup{skip=0pt}
\caption{Offline parameter tuning.}
\label{fig:code_example_a}
\end{subfigure}

\begin{subfigure}{0.95\linewidth}
\centering
\begin{lstlisting}[style=xcodestyle]
engine = KVSWAP.Init( model=model_obj, 
    kv_offload_path="/path/to/offload", 
    config_file="config.json" )
outputs = engine.generate( inputs=[...], 
    sampling_params=[...] )
\end{lstlisting}
\captionsetup{skip=0pt}
\caption{Model serving.}
\label{fig:code_example_b}
\end{subfigure}

\vspace{-2mm}
\caption{Simplified examples for using \DesignName for offline parameter tuning (a) and model serving (b).}
\label{fig:code_example}
\end{figure}

\section{Our Approach}
\SystemName is a Python framework and can support PyTorch-based Transformer LMs, providing simple APIs for model serving and parameter tuning. Figure~\ref{fig:code_example} shows example usage for LM inference: an offline parameter tuning API (Fig.~\ref{fig:code_example}a) generates configurations used by the \SystemName runtime (Fig.~\ref{fig:code_example}b). Using \SystemName is straightforward and typically requires only a few dozen lines of Python code. 


\label{subsec:design_overview}

\begin{figure}[t!]
  \centering
  \captionsetup{skip=0pt}
\includegraphics[width=0.8\columnwidth]{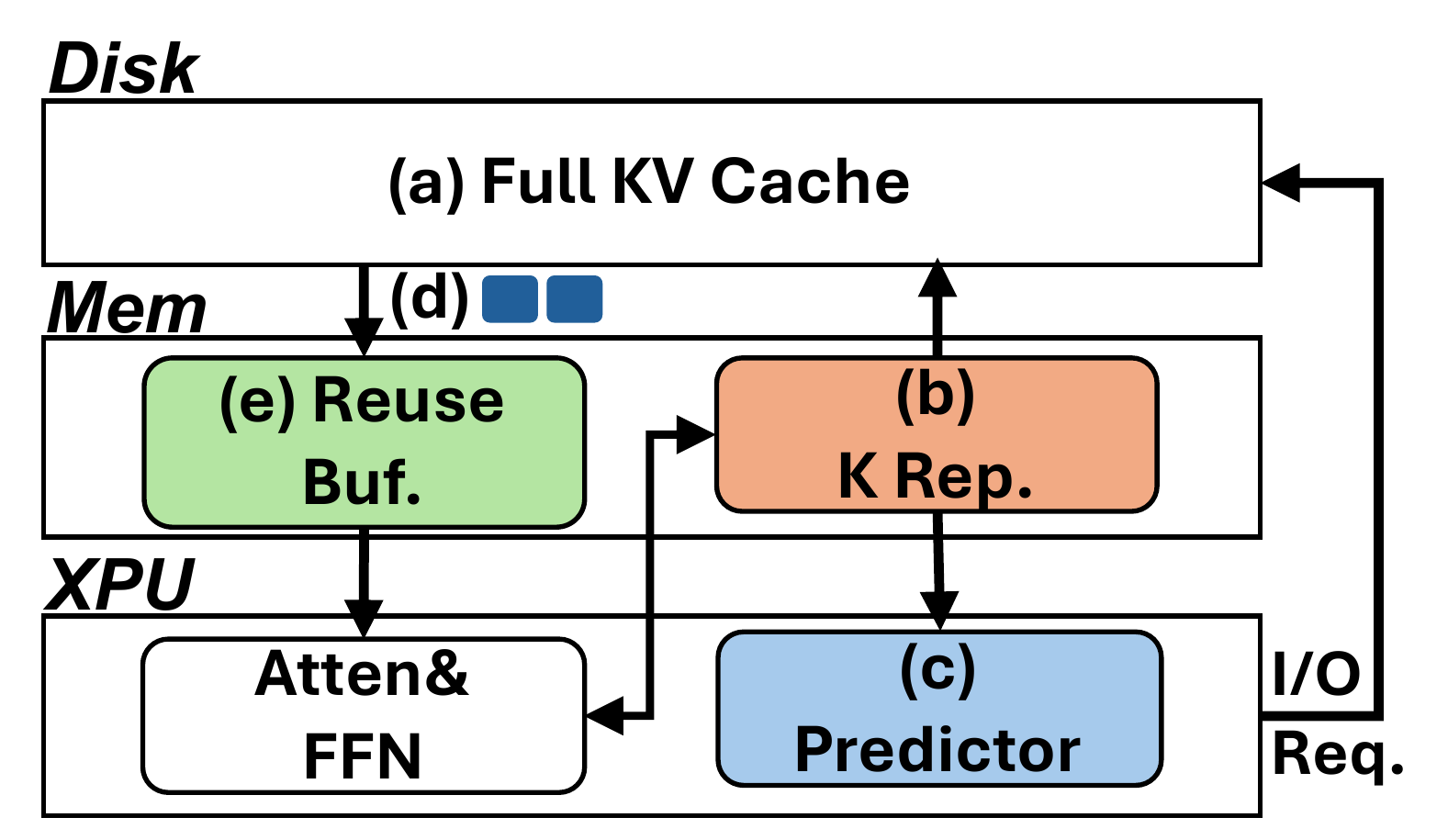}
  \caption{High-level overview of \DesignName.} 
  \label{fig:design}

\end{figure}

Fig.~\ref{fig:design} gives a high-level overview of \DesignName.
\DesignName reduces memory pressure and improves system efficiency by: (a) offloading the full KV cache to disk; (b) maintaining a compact in-memory key cache representation to identify important KV entries using a predictor (c), which will be preloaded from the disk for decoding; (d) minimizing disk I/O overhead by structuring disk access patterns; and (e) effectively caching and reusing previously accessed KV entries through a reuse buffer. 
The \SystemName runtime coordinates I/O and compute, tracks reusable entries across layers, fetches missing ones from disk for the attention kernel, and periodically updates the on-disk full cache and the in-memory representation with newly generated data. 
The KV predictor is detailed in Sec.~\ref{subsec:eff_critical_kv_pred} and Sec.~\ref{subsubsec:group_kv_pred}, the in-memory compression scheme in Sec.~\ref{sec:kcompressed}, and the runtime scheduling and reuse buffer in Sec.~\ref{subsec:runtime_system}.

\subsection{Predicting Important KV Cache Entries\label{subsec:eff_critical_kv_pred}}
The \SystemName KV predictor identifies, \emph{at runtime}, a subset of KV entries needed for each attention computation. Like InfiniGen~\cite{lee2024infinigen}, it estimates the next layer's attention scores from the current layer's input, treating it as an approximate query (Q). These scores quantify the importance of each key and its associated value.
Unlike~\cite{lee2024infinigen}, our setting assumes the entire KV cache resides on disk, where memory and I/O bandwidth are limited. To make prediction feasible, \SystemName maintains a \emph{compact} in-memory representation of the K cache. This compressed cache is used to compute approximate attention scores and select a configurable number of top-ranked KV entries \emph{as groups} for preloading.

Thus, the predictor balances two objectives: (1) minimizing memory footprint, through K cache compression and reconstruction, and (2) improving disk efficiency, through grouped critical KV prediction. These two techniques are detailed in the following subsections.

\subsection{Compressing K Cache\label{sec:kcompressed}}
To enable prediction without an in-memory full K cache, \SystemName maintains a \emph{compressed} K cache in memory using low-rank approximation. Instead of storing every detail of the K matrix, 
we keep a smaller ``summary'' that captures its most important patterns. 
The compressed K cache is automatically managed and updated by the \SystemName runtime  (Sec.~\ref{sec:inmemoryup}). Since the low-rank K cache is used only for estimating attention scores - not for actual attention computation - precision can be flexibly traded for memory efficiency.

We consider two designs:  
(a) \emph{per-head compression}, compressing each head with a separate projection matrix, and 
(b) \emph{joint-head compression}, merging head and embedding dimensions with a single projection.  
We adopt the latter for its unified representation and lower memory cost, reconstructing the multi-head structure during prediction (Sec.~\ref{subsubsec:group_kv_pred}).

We first reshape the K cache into a matrix of size 
$N \times (H_k \cdot d)$, where $N$ is the token count, 
$H_k$ is the number of heads, and $d$ is the head dimension. 
Each vector is then projected into a lower dimension $r$ using a \emph{pre-computed low-rank adapter} 
$A \in \mathbb{R}^{(H_k \cdot d) \times r}$, with $r \ll H_k \cdot d$. 
To obtain $A$, \SystemName applies \emph{Singular Value Decomposition (SVD)}~\cite{eckart1936approximation} 
to a flattened K cache ($K_{\text{ftn}} \in \mathbb{R}^{N \times (H_k \cdot d)}$) 
collected during offline parameter tuning. By default, samples of this K cache are drawn from 
general-purpose datasets~\cite{raffel2020exploring,merity2016pointer}, 
though users may provide their own through the \SystemName API.  Formally, $
    \mathrm{SVD}(K_{\text{ftn}}) = U \, \mathrm{diag}(S) \, V^\top,
$
where $V \in \mathbb{R}^{(H_k \cdot d) \times m}$ and $m = \min(N, H_k \cdot d)$. 
The top-$r$ right singular vectors of $V$ form $A$, and the compressed K cache 
is $K_{\text{lr}} = Flatten(K) A$, with compression level controlled by the compression ratio, $\sigma = \frac{H_k \cdot d}{r}$. We found that the resulting low-rank adapter generalizes well to different datasets.

\cparagraph{Compared with prior K cache compression scheme.} 
While ShadowKV~\cite{sun2024shadowkv} also employs a low-rank K cache, its design and purpose differ from ours. ShadowKV compresses the K cache to reduce its memory footprint, but must reconstruct the full K cache for computation. Hence, it has to use a conservative compression ratio to limit accuracy loss. In contrast, \SystemName uses the low-rank K cache only to predict critical KV entry indices, making it far more resistant to aggressive compression. Combined with our prediction algorithm (Sec.~\ref{subsubsec:group_kv_pred}), this enables \SystemName to maintain generation quality even under high compression ratios, whereas ShadowKV degrades significantly (Sec.~\ref{subsec:gen_quality}).  
The construction method also differs. ShadowKV performs an online SVD at runtime, adding substantial prefill latency (e.g., 4.9$\times$ slower at 8K context length).\footnote{From 7.6s to 37.2s on LLaMA3-8B with NVIDIA Orin AGX, batch size = 1.} \SystemName instead precomputes the low-rank adapter offline, incurring no extra prefill cost.

\subsection{Grouped Critical KV Entries Prediction\label{sec:gckvp}}
\label{subsubsec:group_kv_pred}
\begin{figure}[t!]
  \centering
  \captionsetup{skip=0pt}
  \includegraphics[width=0.95\linewidth]{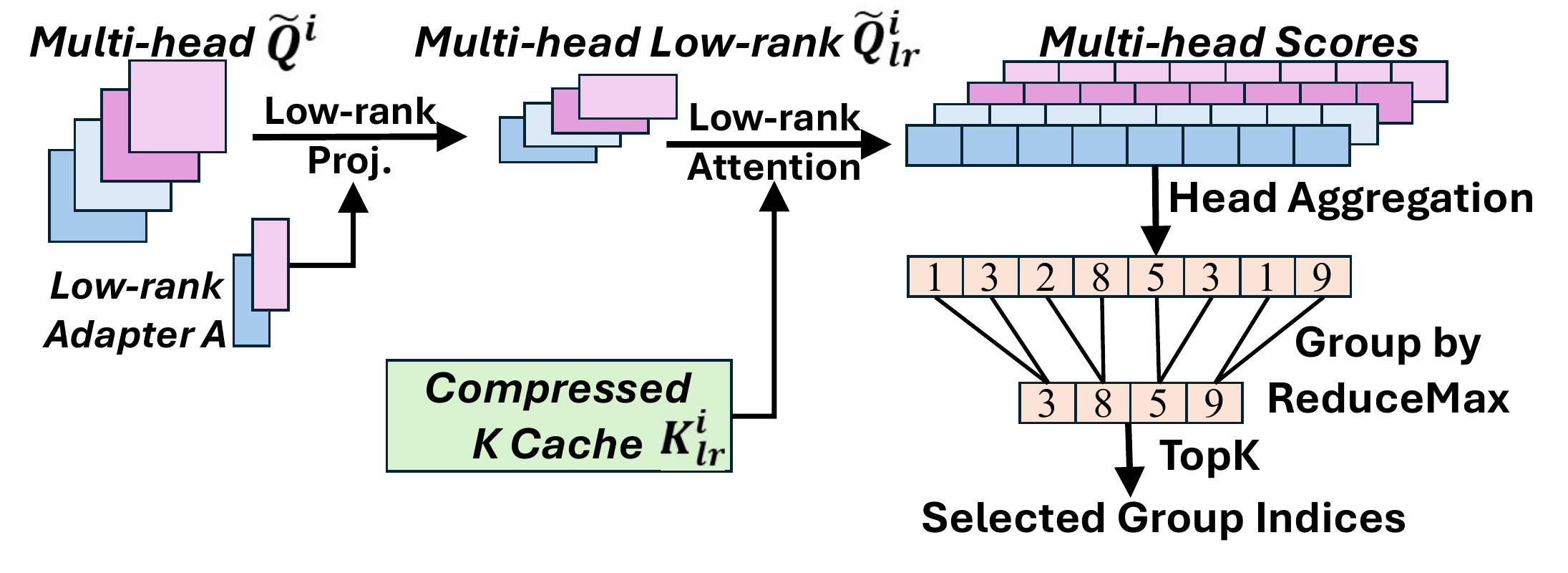}
  \caption{Grouped critical KV entries prediction.}
  \label{fig:prediction}
\end{figure}

To minimize disk I/O,  \SystemName estimates which KV entries are most important for attention computing in the next layers and only preloads these. 
A unique feature of \SystemName is that it first groups G consecutive key-value (KV) entries to align with the block-read characteristics of devices like NVMe, UFS, and eMMC (see also Fig.~\ref{fig:disk_bw_utils}), then predicts and loads the most important groups.
This is different from prior work~\cite{sun2024shadowkv,lee2024infinigen,singhania2024loki} that predicts on individual heads or tokens. The group size is configurable and can be automatically determined during offline processing (Sec.~\ref{subsec:offline_processing}).

Fig.~\ref{fig:prediction} depicts how we predict which \emph{groups} of KV cache entries to be loaded from disk. The idea is to use the low-rank K cache, \(K_{\text{lr}}\), to estimate the attention scores (\(QK^\top\)) without accessing the full K cache. K cache elements (and their associated values, V) that contribute to a high attention score will be prefetched into memory. Specifically, for query head \(h\), we approximate the attention score as: 
\begin{equation}
\small
    Q_h K_{g(h)}^\top \approx Q_h \left( K_{\text{lr}} A_{g(h)}^\top \right)^\top 
    = \left( Q_h A_{g(h)} \right) K_{\text{lr}}^\top
\label{eq:prediction}
\end{equation}
Here, \(Q_h \in \mathbb{R}^{1 \times d}\) is the query vector for head \(h\);  
\(A_{g(h)} \in \mathbb{R}^{d \times r}\) is the low-rank adapter for the K cache assigned to \(h\);  
and \(g(h)\) maps each query head to its shared K head. We refer to the term \(Q_h A_{g(h)}\) as a low-rank query vector. 
This enables the reconstruction of multi-head attention from a compressed, head-unified \(K_{\text{lr}}\), thereby reducing computation and allowing us to decide which \emph{groups} of KV entries are worth loading.

\cparagraph{Online prediction.}  
We predict the critical KV groups ahead of time to issue disk load concurrently with the computation. Specifically, while computing layer \(i{-}1\), \SystemName predicts in advance the critical KV groups needed for layer \(i\).  
To do this, we approximate the upcoming input \(X_i\) using \(X_{i{-}1}\), taking advantage of the observation that the input (embeddings) to a Transformer layer \(i\) is often similar to that at layer \(i{-}1\)~\cite{vaswani2017attention,lee2024infinigen}.
Using this approximation, we apply layer \(i\)’s projections to obtain a low-rank multi-head query vector \(\tilde{Q}_{\text{lr}}^{i}\).  
Together with \(K_{\text{lr}}^{i}\), this vector is used in a low-rank attention computation to estimate the attention scores for layer \(i\).

\cparagraph{Scoring and selection.}  
To select critical KV cache entries, we first compute a single importance score for each token by summing the scores across all heads.
We then form groups of $G$ consecutive tokens and represent each group by the maximum score among its members.
Finally, we select the top-ranked groups with the highest representative scores.
A \texttt{ReduceMax} operation within each group captures the most salient token score, and a \texttt{TopK} step selects the indices of \(M\) most relevant KV entry groups to load for layer \(i\).

\cparagraph{Compared with prior prediction method}.  
Unlike prior work~\cite{lee2024infinigen}, our approach introduces prediction at the group granularity, rather than at individual KV entries. Beyond this, while InfiniGen~\cite{lee2024infinigen}, a representative prior work, also employs approximated attention as its prediction mechanism, our algorithm differs in two fundamental aspects. (1) InfiniGen adopts an index-selecting strategy that pre-determines which embedding dimensions of the K cache to retain and directly selects them via indices. Under stringent K cache compression, this strategy struggles to preserve fidelity as the number of selected indices decreases. (2) InfiniGen applies multi-head attention directly to the index-selected K cache, whereas our method constructs a head-unified low-rank K cache that aggregates information across heads, enabling more aggressive compression. We then reconstruct multi-head attention from this representation using Eq.\ref{eq:prediction}.

\subsection{\SystemName Runtime System}
\label{subsec:runtime_system}

\begin{figure}[!t]
  \centering
  \captionsetup{skip=0pt}
  \includegraphics[width=1.1\linewidth]{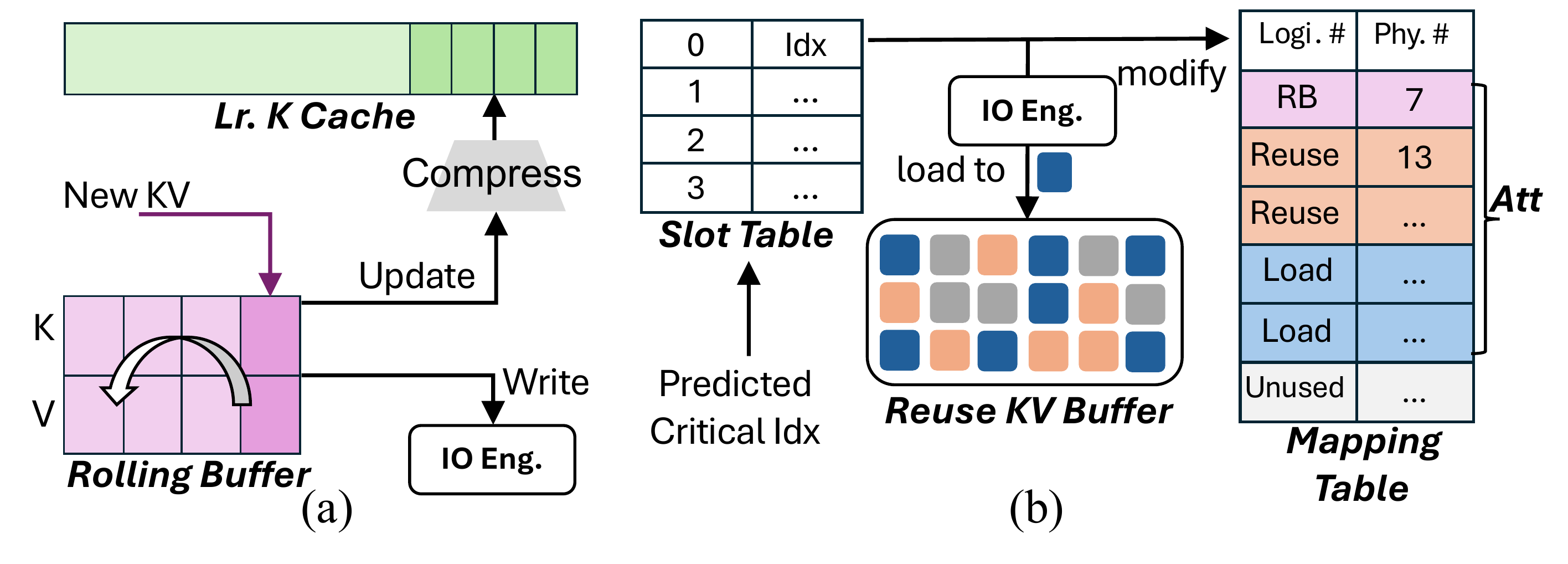}
  \caption{\SystemName runtime: (a) a compressed K cache and (b) reuse buffer for KV cache management.}
  \label{fig:runtime_components}
\end{figure}

\begin{figure}[t!]
  \centering
  \captionsetup{skip=0pt}
  \includegraphics[width=0.95\linewidth,keepaspectratio]{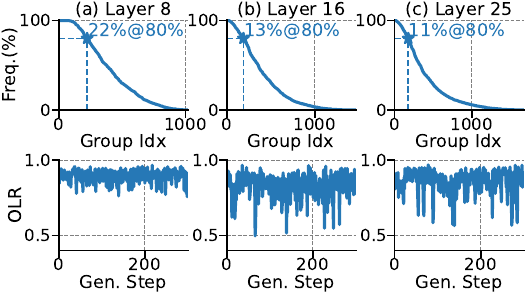}
  \caption{Frequency and overlap ratio (OLR) of predicted critical KV groups over 300 decoding steps. Group indices are reordered by frequency for clarity.}
  \label{fig:gen_overlap}
\end{figure}

We now introduce the overall \SystemName runtime system, which operates during the prefilling and decoding stages (Sec.~\ref{sec:lm}).  
During the \textit{prefilling} phase, \SystemName captures the LM-generated KV cache for the input prompts and writes it to disk in a layer-by-layer fashion. At the same time, it computes a low-rank K cache from the full K cache to create an initial compact K cache (Sec.~\ref{sec:kcompressed}) that is stored in memory.  
When \textit{decoding} begins, the runtime coordinates the grouped critical KV predictor (Sec.~\ref{sec:gckvp}) and the KV cache manager (Sec.~\ref{subsec:kv_cache_manager}) to efficiently support disk-based KV cache loading and attention computation. It identifies, loads, and reuses important KV entry groups while keeping the most recent KV entries in memory. To maximize efficiency and overlap I/O with computation, these operations for the next layer are performed concurrently with the attention and feed-forward network (FFN) computations of the current layer (Sec.~\ref{sec:gckvp}).

\subsubsection{New KV entries management\label{sec:inmemoryup}} 

As shown in Fig.~\ref{fig:runtime_components}a, our in-memory KV cache has two components: a low-rank compressed K cache (Sec.~\ref{sec:kcompressed}) and a rolling buffer (RB). The compressed K cache serves as the input to our KV predictor (Sec.~\ref{sec:gckvp}).  During decoding, new entries are appended to the RB. Since critical entries are predicted in groups, the importance of recent entries cannot be assessed until a group is completed. Thus, the RB temporarily stores recently generated KV entries until they accumulate to a full group of size $G$ for offloading to disk.

\subsubsection{Locality in grouped KV predictions\label{sec:tl}} 
We observe that a subset of critical KV entries exhibits temporal locality across layers. For example, applying our grouped KV predictor (with group size=4) to Qwen3-8B on QMSum~\cite{bai2023longbench} samples, we define the \emph{overlap ratio} at generation step $j$ as the fraction of important groups at step $j$ that also appeared at step $j{-}1$. Fig.~\ref{fig:gen_overlap} reports frequency and overlap ratios over 300 decoding steps\footnote{Results from layers 8, 16, 25 are shown; other layers have similar patterns.}. Fewer than 22\% of groups account for 80\% of occurrences, showing that the set of important groups changes substantially over time. However, adjacent steps exhibit strong overlap, revealing temporal redundancy. This motivates our reuse strategy: retaining recently accessed critical groups in memory allows them to be efficiently reused in subsequent steps without reloading from disk.

\subsubsection{Reuse buffer for caching grouped KV entries.} We introduce a \emph{KV reuse buffer} (not to be confused with the rolling buffer described in Sec.~\ref{sec:inmemoryup} that stores the recently generated KV entries) to store recently accessed critical KV entries (loaded from disk)   for potential reuse across prediction steps. 
By reusing overlapping groups of critical KV entries rather than reloading them, the reuse buffer reduces I/O overhead - especially on low-bandwidth disks or when handling large KV volumes.  We implement the reuse buffer as a fixed set of dedicated \emph{memory slots}, each capable of storing one KV group.  
A slot table records the group ID currently stored in each slot.  
When the predictor identifies the next set of critical groups (Fig.~\ref{fig:runtime_components}b), the system first checks the slot table to see whether a group is already in the reuse buffer.  
If present, the corresponding slot from the reuse buffer is passed to the attention kernel for computation directly; if not, the KV group is loaded from disk and stored in an available slot, with the slot table updated accordingly. We use a simple FIFO policy as the replacement strategy for the reuse buffer.

\subsubsection{KV cache manager\label{subsec:kv_cache_manager}} With \SystemName, attention computation consumes KV entries from: hit slots in the reuse buffer, important KV groups that are not in the reuse buffer and need to be loaded from disk, and recent newly generated entries from the rolling buffer. The \SystemName KV cache manager uses a mapping table to logically address these heterogeneous memory regions - similar to OS virtual memory. This abstraction provides direct compatibility with PagedAttention~\cite{kwon2023efficient}, which expects a contiguous logical view of the KV cache. During generation, reuse patterns change over time, causing the valid memory regions and the positions of usable data in the reuse buffer to shift. To handle this, the runtime updates the mapping table before each attention computation to reflect the current KV entry layout (Fig.~\ref{fig:runtime_components}b).

\subsection{Offline Parameter Tuning} 
\label{subsec:offline_processing}

\SystemName provides an API (Fig.~\ref{fig:code_example}a) for \emph{one-off} offline parameter tuning, which selects optimal runtime settings based on user constraints (maximum memory $\mathcal{B_{\max}}$, context length $S_{\max}$, and batch size $b_{\max}$), the target LM, and the hardware platform. In our evaluation, tuning completes in under 30 minutes and outputs a JSON file that records the best-found parameters for runtime scheduling. These include group size for KV prediction $G$, K-cache compression ratio $\sigma$, number of selected groups $M$, and reuse buffer capacity $\mathcal{C}$ (Sec.~\ref{subsec:runtime_system}).

Parameter search tries to balance three tightly coupled factors: \emph{generation quality}, \emph{inference throughput}, and \emph{memory usage}. For example, $G$, $M$, and $\mathcal{C}$ affect both I/O cost and computation, directly influencing throughput. $G$ and $\sigma$ shape generation quality, while $\sigma$, $M$, and $\mathcal{C}$ determine memory footprint. These interdependencies create a three-way trade-off where improving one dimension often degrades another. Since accurate quality evaluation is also costly, finding a global optimum is impractical. To address this, we use a \emph{greedy-based parameter solver} to efficiently search offline for good local configurations. More details of our parameter search method can be found in Appendix~\ref{details_para_search}.

%% file: sections/5.0.setup.tex
\section{Experimental Setup}
\subsection{Platform and Workloads}

\cparagraph{Evaluation platform.}
We evaluate our approach on the NVIDIA Jetson Orin AGX~\cite{orin} embedded platform with a 12-core Cortex-A78 CPU at 1.3 GHz, an Ampere GPU with 2048 CUDA cores and 64 GB of unified RAM. We apply \SystemName to two widely used disk types on mobile and embedded systems: NVMe (I/O bandwidth 1.8~GB/s) and eMMC (I/O bandwidth 250~MB/s). The system runs JetPack-6.2 with Linux kernel 5.15.148-tegra, CUDA 12.6 and PyTorch 2.7.
\DesignName is designed for scenarios with a tight memory budget. However, as some competing baselines require considerable memory, we leverage the 64 GB of RAM available on the platform to ensure that all methods can be evaluated fairly. Sec.~\ref{subsec:eval_methodology} reports the actual KV-cache memory used in each experimental setting.

\cparagraph{Language models.}
We evaluate \DesignName on both text and video understanding tasks using eight LMs of varying sizes. For text tasks, we use LLaMA-3.1-8B-Instruct (LLaMA3-8B), LLaMA-3.2-3B-Instruct (LLaMA3-3B)~\cite{dubey2024llama}, and Qwen3~\cite{yang2025qwen3} models (4B, 8B, 14B). We enable the thinking mode of Qwen3 to evaluate it on CoT reasoning scenarios. 
For video understanding, we select Qwen2.5-VL-3B, 7B~\cite{bai2025qwen2} and InternVL3-14B~\cite{zhu2025internvl3}.

\cparagraph{Tasks.}
We test on long-context tasks using all English tasks from LongBench~\cite{bai2023longbench}, Needle-in-a-Haystack (NIAH)~\cite{kamradt2023needle}, and eight tasks from RULER~\cite{hsieh2024ruler}. Test context length is fixed at 32K for RULER, and ranges from 4K to 32K for LongBench and NIAH.
For video understanding, we use open-ended generation tasks from MLVU~\cite{zhou2025mlvu}, including Sub-Scene Captioning (SSC) and Video Summarization (VS), with context lengths of 22K--32K per video.

\subsection{Competing Baselines} We compare \DesignName to the following baselines:

\cparagraph{InfiniGen}~\cite{lee2024infinigen}: This closely related work offloads KV cache between GPU and CPU memory by predicting and selecting important KV entries \emph{per} head and token position, unlike our group-based prediction. We extend InfiniGen to the modern GQA attention scheme and adapt its runtime for disk-based offloading. We also evaluate two enhanced variants: \emph{InfiniGen*} adds head aggregation from our prediction strategy (Fig.~\ref{fig:prediction}), and \emph{InfiniGen*+ru} further incorporates our KV reuse strategy to improve disk I/O.

\cparagraph{FlexGen}~\cite{sheng2023flexgen}:  It is designed to offload the entire KV cache to disk. During decoding, the full KV cache is restored layer by layer into memory for full attention computation.

\cparagraph{ShadowKV}~\cite{sun2024shadowkv}: This CPU-GPU KV offloading framework stores a low-rank K cache on the GPU and offloads the V cache to CPU memory (see also Sec.~\ref{sec:kcompressed}). During generation, it loads only important parts of the V cache and reconstructs the K cache on-the-fly. Like InfiniGen, we also adapt its runtime to support disk offloading.

\cparagraph{Loki}~\cite{singhania2024loki}: It is a sparse attention method originally designed to accelerate attention computation. We modify its core approximate attention formulation to function as a predictor for identifying critical KV entries, and embed it into our modified version of InfiniGen with disk-offloading support, where the core prediction components are replaced.

\cparagraph{vLLM}~\cite{kwon2023efficient}: This industry-grade LM serving system stores the full KV cache in memory by default. 
We use vLLM to \emph{approximate the throughput upper bound}, measuring how closely 
an offloading strategy approaches the ideal case without disk overhead. All remaining 
device memory - after storing model weights and allocating workspace - is dedicated to 
the KV cache. We tune vLLM parameters and report the best 
throughput achieved.  

\subsection{Evaluation Methodology\label{subsec:eval_methodology}} 

\cparagraph{Perspective.}
We assess \SystemName under two settings. \textbf{(A)} We align the \emph{per-batch} memory consumption of each method to a consistent, low budget and compare their performance with varying conditions under this constraint. \textbf{(B)} We fix the \emph{total} memory budget and evaluate each method under its best-case configuration. Notably, for baselines at their optimal settings, per-batch memory consumption varies, which directly impacts the maximum achievable batch size for each. 
We evaluate setting A in Sec.~\ref{subsec:gen_quality}, \ref{sec:eval:throughput}, and \ref{subsec:impact_model_sizes}, and setting B in Sec.~\ref{subsec:best_case_eval}.

\cparagraph{Setting A.} In setting A, for all offloading methods, we constrain the \emph{per-batch} KV memory budget of all evaluated LMs to fixed fractions of their corresponding full cache.
We employ two budgets: a relaxed budget ($1/13$ of the full cache) and a tight budget ($1/34$ of the full cache). The latter is denoted with the suffix “-t” (e.g., \DesignName-t). Specifically, we reconfigure the baselines as follows: InfiniGen by adjusting the partial weight ratio~\cite{lee2024infinigen}, Loki by tuning key dimensionality, and ShadowKV by adjusting chunk size, outliers, and K-cache rank.
For \DesignName, we set $S_{\max}=32$K, $b_{\max}=16$, $MG=400$, and maximum K-cache compression ratio $\sigma_{\max}=32$. 

\cparagraph{Setting B.} 
In setting B, we apply the same configuration for \DesignName as in setting A. And for all offloading baselines, we apply their optimal configurations. Specifically, for ShadowKV and LoKi, we directly adopt the configurations reported in their source publications. For InfiniGen, since no reference configurations are readily available, we experimentally set the partial weight ratio to 0.5, resulting in a conservative 4$\times$ KV cache memory reduction. 
Here, their per-batch memory consumption varies and is generally higher than that of \DesignName. To ensure a fair comparison, we instead fix the \emph{total} memory budget and evaluate each method at the \emph{maximum batch size} it can support within this budget. We evaluate under a relaxed total budget of 2000~MiB and a tight total budget of 800~MiB.

\input{sections/exp_data/kvcache_compare}

\cparagraph{Example test memory configuration.}
Tab.~\ref{tab:kv_cache_mem_compare} shows the tested memory budgets for LLaMA3-8B. For vLLM, it is an ideal baseline that uses all available device memory for KV cache (31~GiB) and is not affected by the evaluation settings. Qwen3-8B has a memory budget similar to that of LLaMA3-8B.

\subsection{Performance Metrics} 

We report \emph{generation accuracy} and \emph{throughput} (tokens per second) during decoding. Accuracy is reported under varying KV cache budgets and context lengths. 
Throughput is measured as the average decoding rate over 1000 continuously generated tokens. Each result is the average of five runs with different randomly selected test inputs. Unless otherwise stated, throughput is calculated on LLaMA3-8B, with Qwen3-8B showing similar trends.

%% file: sections/exp_data/kvcache_compare.tex
\begin{table}[!t]
    \centering
    \captionsetup{skip=0pt}
    \footnotesize
    \caption{Tested KV memory budgets for LLaMA3-8B. For \DesignName in both settings (and offloading baselines in setting A), we constrain the \emph{per-batch} KV memory budget of all evaluated LMs to $1/13$ (relaxed) and $1/34$ (tight) of their full KV cache. For setting B, we fix the \emph{total} KV memory budget to 2000~MiB (relaxed) and 800~MiB (tight) for all offloading methods.}
    \label{tab:kv_cache_mem_compare}
    \begin{tabular}{@{} l c c | c@{}}
    \toprule
    \multirow{2}{*}{\textbf{Settings}} & \multicolumn{2}{c|}{\textbf{KV Memory Configuration (MiB)}} & \multirow{2}{*}{\textbf{vLLM (GiB)}} \\
                    & Relaxed & Tight &  \\
    \midrule     
        \textbf{A} (Sec.~\ref{subsec:gen_quality}, \ref{sec:eval:throughput}, \ref{subsec:impact_model_sizes}) & 310/batch@32K  & 120/batch@32K  & \multirow{2}{*}{\centering 31} \\
        \textbf{B} (Sec.~
\ref{subsec:best_case_eval}) & 2000 in total & 800 in total  & \\
    \bottomrule
    \end{tabular}
\end{table}

%% file: sections/5.experiment.tex
\section{Experimental Results\label{sec:results}}

In this section, we show that \SystemName maintains generation quality while delivering strong throughput, outperforming all offloading baselines across diverse settings.

\subsection{LM Generation Quality}
\label{subsec:gen_quality}
\subsubsection{Text processing}
\input{sections/exp_data/main_acc_llama}

\input{sections/exp_data/longbench_cot_vlm}

\input{sections/exp_data/main_tp_16k-32k}

\begin{figure}[t]
  \centering
  \captionsetup{skip=0pt}
  \includegraphics[width=.95\linewidth]{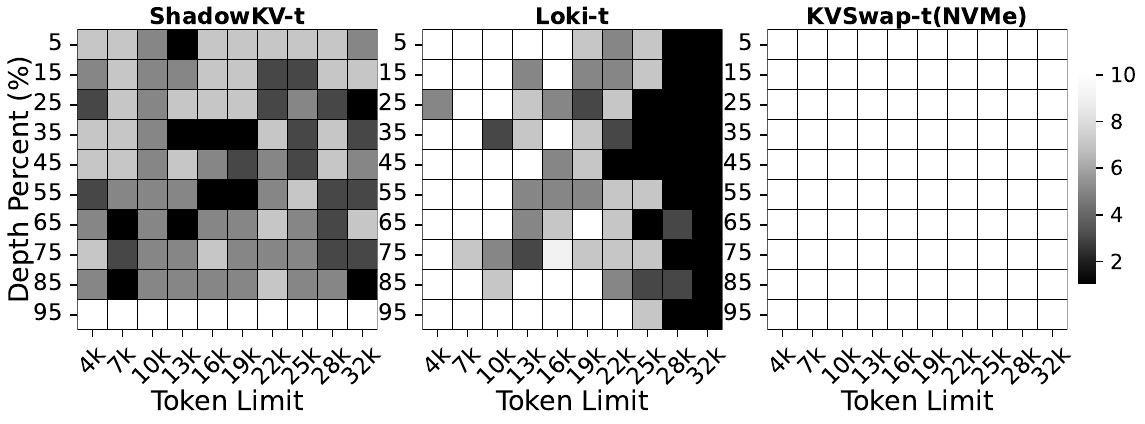}
  \caption{
  Qwen3-8B generation quality on NIAH. x-axis is the tested context length and the y-axis is the relative position to test within a context length.
  }
  \label{fig:needle_results}
\end{figure}

Tab.~\ref{tab:main_acc_llama} reports the generation accuracy of LLaMA3-8B on RULER and LongBench with maximum context length 32K. \emph{Full-KV} shows raw accuracy using the entire KV cache, while other results indicate accuracy loss relative to Full-KV. \SystemName occasionally exceeds Full-KV due to the probabilistic nature of LMs. Results for Qwen3-8B are similar and given in Tab.1 of Appendix~\ref{app:more_eval}.

\DesignName and \DesignName-t (with a tighter memory budget - see Sec.~\ref{subsec:eval_methodology}) outperform all offloading baselines with acceptable accuracy loss.
InfiniGen has the largest degradation, as it fails to capture critical KV entries under tight budgets. Although InfiniGen* reduces accuracy loss by incorporating our head aggregation to reduce prediction noise, it still falls far behind \DesignName, with average accuracy losses of 78.7\% on RULER and 14.2\% on LongBench. 
By contrast, \DesignName keeps average accuracy loss within 4.4\% and 1.1\%. 
ShadowKV leads to 52.3\% and 5.0\% accuracy loss. 
Loki also suffers high losses of 34.0\% and 8.8\%. Furthermore, under a highly constrained budget, only \DesignName-t maintains usable accuracy: $\leq$5.6\% and 2.4\% loss for NVMe. 
Note that \DesignName\textsuperscript{NVMe} and \DesignName\textsuperscript{eMMC} differ because the best group size is $G{=}4$ for NVMe and $G{=}8$ for eMMC.

Fig.~\ref{fig:needle_results} compares \DesignName-t with Loki-t and ShadowKV-t on NIAH with Qwen3-8B under a tight 130~MiB per-batch memory budget (using NVMe). 
The x-axis shows the context length (or token limit) to be tested, and the y-axis shows the relative position to be tested within a context; lower scores with dark regions indicate areas where the model exhibits degraded capability. Only \DesignName-t maintains full processing capability at all sequence locations, while Loki-t and ShadowKV-t suffer substantial generation performance degradation under the same memory budget.

\subsubsection{Reasoning and video understanding}
Tab.~\ref{tab:longbench_cot_vlm} reports generation quality on CoT reasoning and video understanding LMs. For reasoning, we evaluate on three multi-document QA tasks from LongBench (HotpotQA, 2WikiMultihopQA, MuSiQue)~\cite{bai2023longbench}, reporting averaged accuracy across tasks.  
For video understanding, we evaluate Qwen2.5-VL-3B (QVL3B), 7B (QVL7B), and InternVL3-14B (IVL14B), reporting averaged scores on MLVU-SSC and MLVU-VS (range: 2-10).
Recalling evaluation setting A (Sec.~\ref{subsec:eval_methodology}), we configure all offloading methods with per-batch memory budgets set to $1/13$ and $1/34$ of the full KV cache, applied across all models.
We omit InfiniGen and InfiniGen* since Tab.~\ref{tab:main_acc_llama} and Tab.~1 of Appendix~\ref{app:more_eval} already confirm their poor performance. 

Like with text processing, \DesignName outperforms Loki and ShadowKV on NVMe and eMMC disks, with minor accuracy losses $\leq$4.6\%, and score losses $\leq$0.4.  
\DesignName-t also strikes a good balance between KV cache memory footprint and generation quality, while all other methods incur severe degradation in output quality ($\geq$45.0\% accuracy loss on reasoning models and $\geq$2.1 points, or 46.7\%, on video models). 
In contrast, \DesignName-t constraints losses to 3.6-10.0\% on reasoning models and 0.4-1.8 points on video models. Its advantage grows with model size: $\geq$0.3 points on QVL3B, $\geq$1.7 on QVL7B, and $\geq$1.9 on IVL14B.

\cparagraph{Summary.} \DesignName maintains accuracy within an acceptable range across tasks, model sizes, modalities, context lengths, and disk types, consistently outperforming KV cache offloading baselines under the same low-level memory budget. Moreover, \DesignName-t remains robust under a very tight memory budget, while most baselines fail and become impractical for use. 

\begin{figure}[t]
  \centering
  \captionsetup{skip=0pt}
  \includegraphics[width=0.9\linewidth]{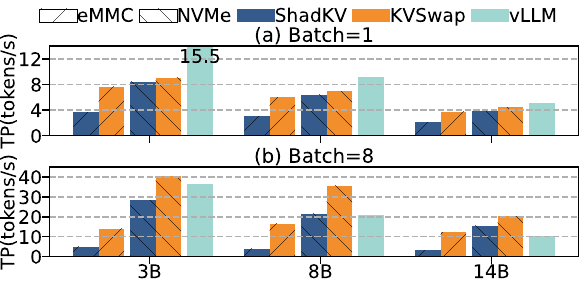}
  \caption{Throughput (TP) comparison across model sizes at a 32K context with batch sizes of (a) 1 and (b) 8.} 
  \label{fig:tp_model_sizes}
\end{figure}

\subsection{LM Generation Throughput}
\label{sec:eval:throughput}
Tab.~\ref{tab:main_throughput_16k_32k} presents generation throughput with batch sizes ranging from 1 to 16 and context lengths of 16K and 32K.
All, except FlexGen and vLLM, use a fixed per-batch KV memory budget of $1/13$ of the full KV cache (see setting A in Tab.~\ref{tab:kv_cache_mem_compare}). FlexGen is unconstrained, as it offloads the full KV cache, while vLLM serves a throughput baseline under idealized memory. Loki and InfiniGen perform similarly due to their fine-grained per-token and per-head design, which yields poor I/O efficiency; their results are averaged for brevity. 
\DesignName outperforms all other offloading methods across storage devices while maintaining the best generation quality (Sec.~\ref{subsec:gen_quality}). Throughput remains stable across context lengths from 16K to 32K since the number of selected KV entries is fixed, making I/O latency largely independent of sequence length.

FlexGen, Loki, and InfiniGen achieve only up to 1.9 and 0.1 tokens/s on NVMe and eMMC respectively due to heavy disk traffic (FlexGen) or poor bandwidth utilization (Loki and InfiniGen). \DesignName sustains 6.9 tokens/s (NVMe) and 5.9 tokens/s (eMMC) with batch size 1 at 16K context, rising to 35.1 and 15.8 tokens/s at batch size 8. On NVMe, throughput continues to scale, reaching 46.1 tokens/s at batch size 16, while eMMC saturates due to its limited bandwidth. Even under saturation, \DesignName maintains significant speedups over competing baselines.
Variants of InfiniGen (InfiniGen* and InfiniGen*+ru) offer moderate improvements but remain far behind \SystemName. For instance, at batch size 4 and 16K context on eMMC, InfiniGen*+ru achieves 5.2 tokens/s versus 16.2 tokens/s for \DesignName, a 3.1$\times$ gain. These results confirm the effectiveness of the grouped KV access pattern we designed.

Finally, \DesignName can sometimes surpass vLLM on NVMe at large batch sizes or contexts by carefully trading LM generation accuracy for throughput. At context length 16K and batch size 16, it is 1.2$\times$ faster while using 11.0$\times$ less KV memory. 
Whereas vLLM saturates once its 31~GiB cache limit is exceeded, \DesignName delivers up to 2.0$\times$ higher throughput with substantially less KV cache memory at an increased 32K context length, a benefit that grows with workload scale.

A complete throughput comparison with additional 8K and 24K context lengths is given in Tab.~2 of Appendix~\ref{app:more_eval}.

\subsection{Impact of LM Model Sizes\label{subsec:impact_model_sizes}}

Fig.~\ref{fig:tp_model_sizes} compares throughput for LLaMA3 (3B and 8B) and Qwen3-14B at a 32K context length with batch sizes of 1 and 8, given available KV cache memory for vLLM of around 43, 31, and 16 GiB. Across model sizes, \DesignName outperforms ShadowKV on eMMC, achieving speedups of over 1.8$\times$ (up to 2.1$\times$) at batch size 1, and over 2.9$\times$ (up to 4.1$\times$) at batch size 8. On NVMe, \DesignName achieves speedups of at least 1.3$\times$ (up to 1.7$\times$) at batch size 8. Although throughput is comparable to ShadowKV on NVMe with batch size 1, \DesignName delivers much better generation quality, as discussed in Sec.~\ref{subsec:gen_quality}.
When compared with vLLM at batch size 8, \DesignName (NVMe) achieves throughput improvements of 1.1$\times$, 1.7$\times$, and 1.9$\times$ on the 3B, 8B, and 14B models, respectively. Notably, on the 14B model, \DesignName with the low-end eMMC disk even surpasses vLLM by 1.2$\times$ at batch size 8.
These results demonstrate that \DesignName scales well as the model grows larger.

\subsection{Best-case Configurations for Baselines} 
\label{subsec:best_case_eval}

\begin{figure}[t]
  \centering
  \captionsetup{skip=0pt}
  \includegraphics[width=0.9\linewidth]{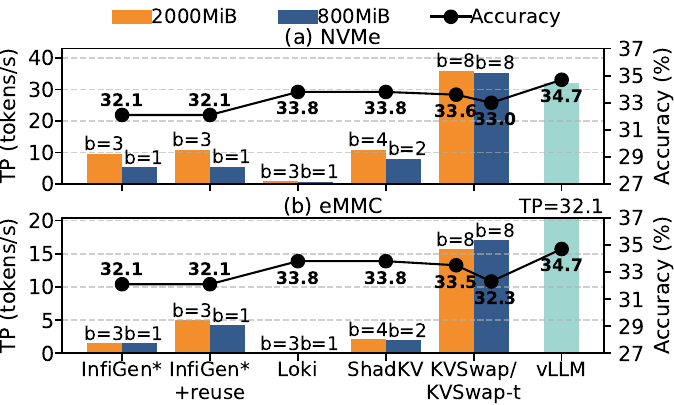}
  \caption{
    Best-case throughput (TP) and accuracy comparison using (a) NVMe and (b) eMMC. $b$ =  largest possible batch size.
  }
  \label{fig:throughput_mem_compare}
\end{figure}

So far, we have set a consistent per-batch memory budget for each KV cache offloading scheme. We now compare \DesignName against offloading baselines using their recommended best-case configurations. Given total memory budgets of 2000~MiB and 800~MiB, we compare throughput using the largest possible batch size supported by each method, as described in Sec.~\ref{subsec:eval_methodology}-setting B.
We also report the average accuracy on two task types of multi-document QA (MQA) and summarization (SUM) from LongBench, six tasks in total, for each method as a reference. 

Fig.~\ref{fig:throughput_mem_compare} presents the results, with vLLM numbers taken from Sec.~\ref{sec:eval:throughput} for direct comparison. \SystemName proves highly effective at trading a small loss in accuracy for substantially higher throughput and much lower memory usage. 
While vLLM, ShadowKV, and Loki achieve the top-3 accuracy, they do so at the cost of either large memory demand or low throughput. In contrast, \DesignName delivers clear advantages in both throughput and memory efficiency, with only marginal accuracy degradation. Compared with vLLM, \DesignName on NVMe reduces KV cache memory by 15.9$\times$–39.7$\times$ while still providing 1.1$\times$ higher throughput, at a maximum accuracy drop of just 2.4\%. Relative to ShadowKV, \DesignName achieves large throughput gains - 7.1$\times$ and 8.6$\times$ on eMMC, and 3.3$\times$ and 4.5$\times$ on NVMe - with accuracy drops $\leq$1.5\%.

%% file: sections/exp_data/main_acc_llama.tex
\begin{table*}[!htbp]
\captionsetup{skip=0pt}
\centering
\caption{Relative accuracy loss (\%) of LLaMA3-8B compared to Full-KV on RULER (left) and LongBench (right).}
\label{tab:main_acc_llama}
\footnotesize
\setlength{\tabcolsep}{1pt}

\begin{tabular}{l|ZZZZZZZZZ|ZZZZZZZ}
\toprule
 \textbf{Methods}
& \textbf{S1} & \textbf{S2} & \textbf{MK1} & \textbf{MQ} & \textbf{MV}
& \textbf{QA1} & \textbf{QA2} & \textbf{VT} & \textbf{Avg.}
& \textbf{SQA} & \textbf{MQA} & \textbf{SUM} & \textbf{FSL} & \textbf{SYN} & \textbf{COD} & \textbf{Avg.} \\
\midrule

\cellcolor{lightblue}Full-KV (raw \%)& \cellcolor{lightblue}100.0 & \cellcolor{lightblue}100.0 & \cellcolor{lightblue}100.0 & \cellcolor{lightblue}98.75 & \cellcolor{lightblue}99.5 & \cellcolor{lightblue}82.0 & \cellcolor{lightblue}56.0 & \cellcolor{lightblue}96.61 & \cellcolor{lightblue}91.6 & \cellcolor{lightblue}39.33 & \cellcolor{lightblue}41.39 & \cellcolor{lightblue}28.03 & \cellcolor{lightblue}67.99 & \cellcolor{lightblue}99.5 & \cellcolor{lightblue}49.08 & \cellcolor{lightblue}54.22 \\
\cline{1-17}
InfiniGen   & -100.0 & -100.0 & -100.0 & -98.75 & -99.5 & -61.0 & -40.0 & -96.61 & -86.98 & -36.01 & -37.75 & -26.82 & -50.8  & -99.0  & -44.7  & -49.18 \\
InfiniGen*  & -86.0  & -97.0  & -97.0  & -98.25 & -99.5 & -39.0 & -20.0 & -93.01 & -78.71 & -14.37 & -14.81 & -10.69 & -14.58 & -5.25  & -25.42 & -14.19 \\
Loki        & -3.0   & -14.0  & -18.0  & -67.5  & -61.5 & -28.0 & -15.0 & -65.41 & -34.04 & -7.39  & -4.86  & -10.78 & -7.96  & -0.3  & -21.64 & -8.81 \\
ShadowKV    & \bfseries0.0 & -86.0  & -84.0  & -93.0  & -94.5 & -13.0 & -7.0  & -41.01 & -52.31 & -9.73  & -2.8  & -4.1  & -5.07  & -1.5   & -6.8  & -4.99 \\
\cellcolor{lightgreen}\DesignName\textsuperscript{NVMe} & \cellcolor{lightgreen}\bfseries 0.0 & \cellcolor{lightgreen}-1.0 & \cellcolor{lightgreen}\bfseries0.0 & \cellcolor{lightgreen}\bfseries-3.3 & \cellcolor{lightgreen}\bfseries-7.0 & \cellcolor{lightgreen}-1.0 & \cellcolor{lightgreen}\bfseries{+2.0} & \cellcolor{lightgreen}\bfseries-10.4 & \cellcolor{lightgreen}\bfseries-2.6 & \cellcolor{lightgreen}\bfseries{+0.9} & \cellcolor{lightgreen}-0.26 & \cellcolor{lightgreen}\bfseries-2.05 & \cellcolor{lightgreen}\bfseries-0.74 & \cellcolor{lightgreen}\bfseries0.0 & \cellcolor{lightgreen}\bfseries-1.8 & \cellcolor{lightgreen}\bfseries-0.6 \\
\cellcolor{lightgreen}\DesignName\textsuperscript{eMMC} & \cellcolor{lightgreen}\bfseries0.0 & \cellcolor{lightgreen}\bfseries0.0 & \cellcolor{lightgreen}\bfseries0.0 & \cellcolor{lightgreen}-9.0 & \cellcolor{lightgreen}-12.0 & \cellcolor{lightgreen}\bfseries0.0 & \cellcolor{lightgreen}-3.0 & \cellcolor{lightgreen}-11.8 & \cellcolor{lightgreen}-4.4 & \cellcolor{lightgreen}-0.5 & \cellcolor{lightgreen}\bfseries-0.1 & \cellcolor{lightgreen}-2.3 & \cellcolor{lightgreen}-1.4 & \cellcolor{lightgreen}-0.5 & \cellcolor{lightgreen}\bfseries-1.8 & \cellcolor{lightgreen}-1.1 \\
\cline{1-17}
Loki-t      & -99.0  & -100.0 & -100.0 & -98.75 & -99.5 & -57.0 & -32.0 & -96.61 & -85.35 & -24.45 & -25.67 & -20.46 & -43.2  & -7.59  & -38.2  & -26.6 \\
ShadowKV-t  & -21.0  & -89.0  & -93.0  & -95.75 & -95.0 & -16.0 & -6.0  & -79.21 & -61.86 & -15.45 & -7.0   & -7.1   & -10.18 & -4.5   & -7.96  & -8.7 \\
\cellcolor{lightgreen}\DesignName-t\textsuperscript{NVMe} & \cellcolor{lightgreen}\bfseries0.0 & \cellcolor{lightgreen}\bfseries-1.0 & \cellcolor{lightgreen}\bfseries-3.0 & \cellcolor{lightgreen}\bfseries-10.8 & \cellcolor{lightgreen}\bfseries-13.3 & \cellcolor{lightgreen}\bfseries-4.0 & \cellcolor{lightgreen}\bfseries0.0 & \cellcolor{lightgreen}\bfseries-13.0 & \cellcolor{lightgreen}\bfseries-5.6 & \cellcolor{lightgreen}\bfseries-2.5 & \cellcolor{lightgreen}\bfseries-0.4 & \cellcolor{lightgreen}\bfseries-3.0 & \cellcolor{lightgreen}-2.7 & \cellcolor{lightgreen}\bfseries0.0 & \cellcolor{lightgreen}-5.8 & \cellcolor{lightgreen}\bfseries-2.4 \\
\cellcolor{lightgreen}\DesignName-t\textsuperscript{eMMC} & \cellcolor{lightgreen}\bfseries0.0 & \cellcolor{lightgreen}-13.0 & \cellcolor{lightgreen}-16.0 & \cellcolor{lightgreen}-54.3 & \cellcolor{lightgreen}-55.8 & \cellcolor{lightgreen}-13.0 & \cellcolor{lightgreen}-7.0 & \cellcolor{lightgreen}-24.6 & \cellcolor{lightgreen}-22.9 & \cellcolor{lightgreen}-4.2 & \cellcolor{lightgreen}-1.1 & \cellcolor{lightgreen}-3.9 & \cellcolor{lightgreen}\bfseries-1.5 & \cellcolor{lightgreen}-0.5 & \cellcolor{lightgreen}\bfseries-5.2 & \cellcolor{lightgreen}-2.7 \\
\bottomrule
\end{tabular}
\end{table*}

%% file: sections/exp_data/longbench_cot_vlm.tex
\begin{table}[!htbp]
    \centering
    \captionsetup{skip=0pt}
    \footnotesize
    \setlength{\tabcolsep}{1.5pt}
    \caption{
        Relative accuracy (\%) and score loss over Full-KV on reasoning (left) and video understanding (right).
    }
    \label{tab:longbench_cot_vlm}
    \begin{tabular}{@{}lZZZ|ZZZ@{}}
        \toprule
        \textbf{Methods} & \textbf{Q4B} & \textbf{Q8B} & \textbf{Q14B} & \textbf{QVL3B} & \textbf{QVL7B} & \textbf{IVL14B} \\
        \midrule
        \cellcolor{lightblue}Full-KV (raw \%) & \cellcolor{lightblue}60.47 &  \cellcolor{lightblue}62.47  & \cellcolor{lightblue}65.98 & \cellcolor{lightblue}4.52 & \cellcolor{lightblue}4.87 & \cellcolor{lightblue}4.76\\
        \midrule
        Loki          & -10.02 & -4.06  & -9.56 & -1.36 & -1.02 & -2.01\\
        ShadowKV      & -6.62 & -5.37 & -7.34 & -0.71 & -0.55 & -0.53 \\
        \cellcolor{lightgreen}\DesignName\textsuperscript{NVMe}   & \cellcolor{lightgreen}\bfseries-1.8 & \cellcolor{lightgreen}\bfseries-2.4 & \cellcolor{lightgreen}\bfseries-2.6 & \cellcolor{lightgreen}\bfseries-0.4 & \cellcolor{lightgreen}\bfseries-0.3 & \cellcolor{lightgreen}\bfseries-0.3 \\
        \cellcolor{lightgreen}\DesignName\textsuperscript{eMMC}   & \cellcolor{lightgreen}-4.0 & \cellcolor{lightgreen}-3.0 & \cellcolor{lightgreen}-4.6 & \cellcolor{lightgreen}\bfseries-0.4 & \cellcolor{lightgreen}\bfseries-0.3 & \cellcolor{lightgreen}\bfseries-0.3 \\
        \midrule
        Loki-t        & -57.20 & -47.24  & -58.63   & -2.52 & -2.87 & -2.74\\
        ShadowKV-t    & -50.10 & -47.51 & -45.01 & -2.14 & -2.36 & -2.44 \\
        \cellcolor{lightgreen}\DesignName-t\textsuperscript{NVMe} & \cellcolor{lightgreen}\bfseries-3.6 & \cellcolor{lightgreen}\bfseries-4.8 & \cellcolor{lightgreen}\bfseries-6.7 & \cellcolor{lightgreen}\bfseries-1.7 & \cellcolor{lightgreen}\bfseries-0.6 & \cellcolor{lightgreen}\bfseries-0.4\\
        \cellcolor{lightgreen}\DesignName-t\textsuperscript{eMMC} & \cellcolor{lightgreen}-7.2 & \cellcolor{lightgreen}-6.9 & \cellcolor{lightgreen}-10.0 & \cellcolor{lightgreen}-1.8 & \cellcolor{lightgreen}-0.7 & \cellcolor{lightgreen}-0.5 \\
        \bottomrule
    \end{tabular}
    \vspace{-2mm}
\end{table}

%% file: sections/exp_data/main_tp_16k-32k.tex
\begin{table}[!htbp]
\captionsetup{skip=0pt}
\centering
\caption{Throughput (tokens/s) comparison of LLaMA3-8B across batch sizes and context lengths (CLs).
}
\label{tab:main_throughput_16k_32k}
\footnotesize
\setlength{\tabcolsep}{0.55pt}
\begin{tabular}{l|l|SSSSS|SSSSS}
\toprule
 & \multirow{2}{*}{\textbf{Methods}} 
& \multicolumn{5}{c}{\textbf{CL=16K}} 
& \multicolumn{5}{c}{\textbf{CL=32K}} \\
\cmidrule(lr){3-7} \cmidrule(lr){8-12} 
& & \textbf{b=1} & \textbf{b=2} & \textbf{b=4} & \textbf{b=8} & \textbf{b=16}
  & \textbf{b=1} & \textbf{b=2} & \textbf{b=4} & \textbf{b=8} & \textbf{b=16} \\
\midrule
\multirow{7}*[-0.4ex]{\rotatebox[origin=c]{90}{\makebox[1.2cm][r]{\strut eMMC}}}
 & \cellcolor{gray!20}FlexGen &\cellcolor{gray!20}0.11 &\cellcolor{gray!20}0.1 &\cellcolor{gray!20}0.11 &\cellcolor{gray!20}0.11 &\cellcolor{gray!20}0.11  &\cellcolor{gray!20}0.05 &\cellcolor{gray!20}0.05 &\cellcolor{gray!20}0.05 &\cellcolor{gray!20}0.05 &\cellcolor{gray!20}0.05 \\
 & Loki/InfiGen& 0.1 & 0.1 & 0.1 & 0.1 & 0.1 & 0.1 & 0.1 & 0.1 & 0.1 & 0.1\\
 & \cellcolor{gray!20}InfiGen* & \cellcolor{gray!20}1.71 & \cellcolor{gray!20}1.65 & \cellcolor{gray!20}1.46 & \cellcolor{gray!20}1.44 &\cellcolor{gray!20}1.40 & \cellcolor{gray!20}1.61 & \cellcolor{gray!20}1.51 & \cellcolor{gray!20}1.38 & \cellcolor{gray!20}1.35 & \cellcolor{gray!20}1.35\\
 & InfiGen*+ru & 3.96 & 4.69 &5.24 &4.23 &3.71 & 4.2 & 4.4 & 4.9 & 4.5 & 3.5\\
 & \cellcolor{gray!20}ShadowKV & \cellcolor{gray!20}2.97 & \cellcolor{gray!20}3.76 & \cellcolor{gray!20}4.14 & \cellcolor{gray!20}4.40 & \cellcolor{gray!20}3.44 & \cellcolor{gray!20}2.97 & \cellcolor{gray!20}3.72 & \cellcolor{gray!20}4.17 & \cellcolor{gray!20}3.76 & \cellcolor{gray!20}3.4 \\
 & \cellcolor{lightgreen}\DesignName &\cellcolor{lightgreen}\bfseries5.94 &\cellcolor{lightgreen}\bfseries10.45 &\cellcolor{lightgreen}\bfseries16.16 &\cellcolor{lightgreen}\bfseries15.79 &\cellcolor{lightgreen}\bfseries11.18  &\cellcolor{lightgreen} \bfseries5.96&\cellcolor{lightgreen}\bfseries10.25  &\cellcolor{lightgreen}\bfseries15.19 &\cellcolor{lightgreen}\bfseries15.69 &\cellcolor{lightgreen}\bfseries10.9 \\
\midrule

\multirow{7}*[-0.4ex]{\rotatebox[origin=c]{90}{\makebox[1.2cm][r]{\strut NVMe}}}
 & \cellcolor{gray!20}FlexGen  &\cellcolor{gray!20}0.8 &\cellcolor{gray!20}0.8 &\cellcolor{gray!20}0.8 &\cellcolor{gray!20}0.8 &\cellcolor{gray!20}0.8  &\cellcolor{gray!20}0.4 &\cellcolor{gray!20}0.4 &\cellcolor{gray!20}0.4 &\cellcolor{gray!20}0.4 &\cellcolor{gray!20}0.4 \\
 & Loki/InfiGen  & 1.9 & 1.9 & 1.9 & 1.9 & 1.9 & 1.9 & 1.9 & 1.9 & 1.8&1.8 \\
 & \cellcolor{gray!20}InfiGen* & \cellcolor{gray!20}4.95 & \cellcolor{gray!20}8.36 & \cellcolor{gray!20}10.81 & \cellcolor{gray!20}11.32 & \cellcolor{gray!20}13.1  &\cellcolor{gray!20}5.1 &\cellcolor{gray!20}7.52 &\cellcolor{gray!20}8.88 &\cellcolor{gray!20}10.71 &\cellcolor{gray!20}11.96 \\
 & InfiGen*+ru  & 5.22 & 9.15 & 14.86 & 22.50 &26.57 & 4.93 & 8.5 & 11.89 & 14.33 & 17.22 \\
 &\cellcolor{gray!20}ShadowKV & \cellcolor{gray!20}6.41 &\cellcolor{gray!20} 10.43 &\cellcolor{gray!20} 16.31 &\cellcolor{gray!20} 21.85 &\cellcolor{gray!20} 26.71  &\cellcolor{gray!20} 6.38 &\cellcolor{gray!20} 10.03 &\cellcolor{gray!20} 16.25 &\cellcolor{gray!20} 21.49 &\cellcolor{gray!20} 26.21\\
 
 & \cellcolor{lightgreen}\DesignName &\cellcolor{lightgreen}\bfseries6.93 &\cellcolor{lightgreen}\bfseries11.91 &\cellcolor{lightgreen}\bfseries20.80 &\cellcolor{lightgreen}\bfseries35.05 &\cellcolor{lightgreen}\bfseries46.06  &\cellcolor{lightgreen}\bfseries6.85 &\cellcolor{lightgreen}\bfseries11.9 &\cellcolor{lightgreen}\bfseries21.41 &\cellcolor{lightgreen}\bfseries35.6 &\cellcolor{lightgreen}\bfseries46.79 \\

\midrule

- & \cellcolor{lightblue}vLLM &\cellcolor{lightblue} 9.69 &\cellcolor{lightblue} 17.14 &\cellcolor{lightblue} 28.38 &\cellcolor{lightblue} 41.19 &\cellcolor{lightblue} 39.46 &\cellcolor{lightblue} 9.22 &\cellcolor{lightblue} 14.06 &\cellcolor{lightblue} 21.03 &\cellcolor{lightblue} 20.77 &\cellcolor{lightblue} 23.21 \\
\bottomrule
\end{tabular}

\end{table}

%% file: sections/6.analysis.tex
\section{Analysis of Design Choices}
\label{sec:analysis}
To evaluate how our key design choices, including group size, KV reuse buffer and number of selected KV entries, affect the overall throughput and accuracy, we apply \DesignName to LLaMA3-8B under a 310 MiB per-batch KV memory budget. Accuracy is reported as the average over MV, QA1, and VT tasks from RULER, while throughput and latency are measured at a 32K context length with a batch size of 8.

\begin{figure}[t]   
  \centering
  \captionsetup{skip=0pt}
  \includegraphics[width=0.9\linewidth]{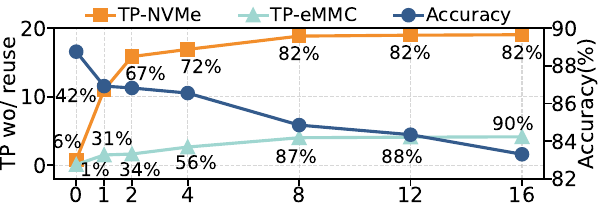}
    \caption{
    Trade-off between accuracy, I/O efficiency, and throughput (TP) under varying group sizes (x-axis). 
    }
  \label{fig:ablations_group}
\end{figure}

\input{sections/exp_data/reuse_buf}

\cparagraph{Group size and system efficiency}. 
Figure~\ref{fig:ablations_group} shows how throughput and accuracy vary with KV prediction group sizes ($G$). The number beside each point indicates I/O utilization. 
A group size of 0 disables our head aggregation strategy. This ablation study reports throughput \emph{without} KV reuse to isolate our optimizations. As $G$ increases, accuracy drops gradually from $88.8\%$ to $83.3\%$, while throughput (without reuse) improves from 1.8 to 19.1 tokens/s on NVMe and 0.1 to 4.2 tokens/s on eMMC. 
When $G=0$ or $1$, both throughput and I/O utilization are low. 
This highlights that our grouped critical KV prediction is important for improving the throughput while maintaining generation accuracy.
\SystemName provides an API to automatically obtain the best group size, but users can configure \SystemName to prioritize generation accuracy or throughput.

\cparagraph{KV reuse and throughput.} To analyse the characteristics and benefits of KV reuse, we evaluate 100 randomly sampled inputs: 50 from the QMSum meeting summarization dataset and 50 from the MuSiQue multi-document QA dataset, covering different use scenarios. 
Tab.~\ref{tab:reuse_buf} reports reuse rates and throughput across disk types and datasets. Reuse rates remain high across all cases, ranging from 75.3\% to 81.2\%.
The throughput (in tokens/s) achieves $2.0\times$–$2.1\times$ speedups on NVMe and $3.8\times$–$4.0\times$ on eMMC compared to the no-reuse baseline.
This indicates that our reuse buffer can effectively improve the throughput, especially for slower persistent storage.
Similar trends hold across the evaluated workloads: reuse rates vary little across inputs (standard deviation $\leq 1.1\%$), 
and throughput variation remains bounded within 2.0 standard deviations on NVMe and 0.6 on eMMC, contributing to no more than 5.6\% of the average throughput. These results justify using the average reuse rate during offline tuning (Appendix~\ref{sec:lookup}).

\begin{figure}[t]
  \centering
  \captionsetup{skip=0pt}
  \includegraphics[width=\linewidth]{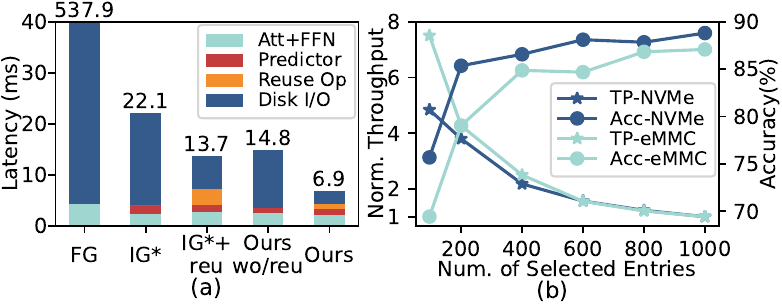}
  \caption{
    (a) Decoding latency breakdown on NVMe. 
    (b) Trade-off between accuracy and throughput across KV selection sizes and disk types.
  }
  \label{fig:ablation_m_latency_breakdown}
\end{figure}

\cparagraph{Latency breakdown.}
Figure~\ref{fig:ablation_m_latency_breakdown}a reports the decoding latency of a single Transformer block. In FlexGen (FG), loading all KV entries from disk makes I/O the bottleneck. InfiniGen* (IG*) alleviates this with selective loading and our head aggregation scheme, but I/O latency still dominates.
\SystemName without reuse (Ours wo/ reu) better utilizes disk bandwidth, reducing latency by 1.5$\times$, already comparable to InfiniGen*+reuse (IG*+reu). With reuse enabled, I/O latency of \SystemName further drops by 4.3$\times$, incurring only 1.0~ms reuse overhead, and achieves the lowest total latency of 6.9~ms.

\cparagraph{Number of selected entries.}
Figure~\ref{fig:ablation_m_latency_breakdown}b shows normalized throughput versus the number of selected KV entries $MG$ (omitting the number of KV heads $H_{kv}$).
Increasing $MG$ improves accuracy on NVMe and eMMC but reduces throughput. Beyond $MG=400$, accuracy gains become marginal while throughput continues to drop, making $MG=400$ the balanced trade-off and our default setting.

\cparagraph{Rolling buffer.} We compare generation accuracy with and without the rolling buffer (RB) on KV entry group sizes ($G$) 2, 4, 8 and 12 (see also Tab.~3 of Appendix ~\ref{app:more_eval}). Disabling RB causes a sharp accuracy drop of  $\ge$29\%, since new KV entries are generated token by token during decoding and often cannot immediately form a group for prediction. The RB allows newly generated entries to participate in computation in a timely manner, helping to preserve accuracy.

\section{Discussion}
\label{sec:discussion}

\vspace{-0.5mm}
\cparagraph{Model weight optimization.} \SystemName optimizes KV cache memory footprint. Weight-centric methods such as quantization~\cite{lin2024awq,frantar2022gptq} and parameter offloading~\cite{alizadeh2023llm,xue2024powerinfer} are orthogonal and complementary.

\cparagraph{Low-bit KV.} \SystemName is designed to preserve generation quality, but can be combined with low-bit KV compression~\cite{liu2024kivi,lin2024qserve}. For comparison, 2-bit quantization yields an $8\times$ reduction, while \SystemName achieves more than $30\times$.

\cparagraph{Prefilling optimization.} \SystemName can extend to prefilling by (1) computing attention only for important tokens, and (2) storing only critical KV entries to disk. It also complements prefill-oriented frameworks~\cite{jiang2024minference}.

\cparagraph{OS paging.} Using OS-level paging is inefficient, as it is workload-agnostic and misses optimizations like selective KV loading, reduced transfer, and overlapped prefetching.

\cparagraph{ShadowKV+reuse.} While employing reuse to ShadowKV~\cite{sun2024shadowkv} may improve throughput, it cannot meet the tight on-device memory limits. By algorithm and system co-design, \DesignName improves both memory efficiency and system throughput.

%% file: sections/exp_data/reuse_buf.tex
\begin{table}[t]
    \centering
    \captionsetup{skip=0pt}
    \footnotesize
    \setlength{\tabcolsep}{3pt}
    \caption{Reuse ratio and throughput (TP) statistics. "No Reu." = TP wo/ reuse. "$\uparrow$" = TP improvement of w/ over wo/ reuse.}
    \label{tab:reuse_buf}
    \begin{tabular}{llrrrrrcr}
        \toprule
        \textbf{Disk} & \textbf{Dataset} & \textbf{Metric} & \textbf{Min} & \textbf{Max} & \textbf{Std} & \textbf{Avg} & \textbf{No Reu.} & \textbf{$\uparrow$} \\
        \midrule
        \multirow{4}{*}{NVMe}
          & \multirow{2}{*}{QMSum} & Reuse(\%) & 76.2 & 79.1 & 0.7 & 77.3 &  - & \multirow{2}{*}{2.1$\times$} \\
          &                        & TP(toks/s)    & 33.4 & 38.7 & 2.0 & 35.8 & 17.3     &                                      \\
          \cline{2-9}
          & \multirow{2}{*}{MSQue} & Reuse(\%) & 75.3 & 81.2 & 1.1 & 76.2 &  - & \multirow{2}{*}{2.0$\times$} \\
          &                   & TP(toks/s) & 32.4 & 37.6 & 1.5 & 35.4 & 17.4  &                        \\
        \midrule

        \multirow{4}{*}{eMMC}
          & \multirow{2}{*}{QMSum} & Reuse(\%) & 76.1 & 79.2 & 0.7 & 77.2 & - & \multirow{2}{*}{4.0$\times$} \\
          &                        & TP(toks/s)    & 14.4 & 17.5 & 0.6 & 15.7 &  3.9   &                        \\
          \cline{2-9}
          & \multirow{2}{*}{MSQue} & Reuse(\%) & 76.3 & 79.4 & 0.7 & 77.6 & - & \multirow{2}{*}{3.8$\times$} \\
          &                        & TP(toks/s)    & 14.5 & 16.0 & 0.4 & 15.3 & 4.0 &  \\
        \bottomrule
    \end{tabular}

\end{table}

%% file: sections/7.related_work.tex
\section{Related Work}
\SystemName builds on the following past foundations.

\cparagraph{Sparse attention.}
Sparse attention~\cite{rewon2019sparse} reduces the memory and compute overhead of 
attention and KV cache. Static methods apply fixed patterns to select KV entries~\cite{Iz2020longformer,bigbird2020,ding2023longnet,jin2024icml}, 
reducing computation but risking loss of long-range dependencies and accuracy. 
Dynamic methods adapt token selection to the input, focusing on tokens that dominate 
attention scores~\cite{reformer2020iclr,zhang2023h2o,SpAtten2021hpca,dynamic2023nips,zhang2024semsa,singhania2024loki,wu2025tokenselect,quest2024icml,zhang2021sparse,gated2024JMLR,yuan-etal-2025-native}.  
\SystemName builds on this insight by identifying important KV entries 
at each step, but adapts it for on-device disk offloading. Unlike sparse attention, which assumes KV cache is fully in memory, \SystemName selectively preloads critical KV groups 
from disk into memory using two techniques: a compressed K cache 
(Sec.~\ref{sec:kcompressed}) to reduce memory overhead, and a grouped KV prediction 
algorithm (Sec.~\ref{subsubsec:group_kv_pred}) to improve disk I/O.

\cparagraph{KV-cache offloading and optimization.}
InfiniGen~\cite{lee2024infinigen} and ShadowKV~\cite{sun2024shadowkv} move KV entries between GPU and CPU, while SpeCache~\cite{jie2025specache} applies low-bit compression and speculative prefetching. llm-offload~\cite{llmoffload2025IPDPSW} and ArkVale~\cite{arkvalue2024nips} adopt page-based KV managers to recall evicted tokens. These \emph{GPU-CPU offloading} methods assume high-bandwidth, low-latency PCIe connections, whereas mobile and embedded devices rely on NVMe, UFS, eMMC, or SD storages with orders-of-magnitude lower bandwidth and higher latency. This makes disk-based offloading a fundamentally harder problem. \SystemName is the first framework explicitly designed to tackle this challenge.
Optimizations are also proposed to \textit{time-to-first-token (TTFT)} to reduce startup delay. CacheBlend fuses cached knowledge for faster retrieval-augmented generation~\cite{yao2025cacheblend}, IMPRESS~\cite{chen2025impress} accelerates prefix KV loading on server-class SSDs, and CacheGen~\cite{liu2024cachegen} compresses KV for faster transmission. In contrast, \SystemName targets iterative decoding on resource-constrained devices, where decoding throughput - not TTFT - is the primary bottleneck.

\cparagraph{On-device inference.}
General-purpose LM serving frameworks such as llama.cpp~\cite{llama_cpp} and MLC-LLM~\cite{mlc_llm} enable inference on consumer hardware. Beyond these, recent work has explored memory-efficient serving through quantization and weight optimization. Other approaches exploit contextual sparsity~\cite{liu2023deja}, weight offloading~\cite{alizadeh2023llm,xue2024powerinfer}, or model specialization such as EdgeMoE~\cite{yi2023edgemoe}. System-level advances include LLMCad~\cite{xu2023llmcad} for speculative decoding and architectural redesigns for mobile platforms~\cite{team2024gemma, liu2024mobilellm, mehta2024openelm, thawakar2024mobillama}.
Most of these efforts focus on optimizing \emph{model weights}. In contrast, \SystemName tackles the growing overhead of the \emph{KV cache}, the main bottleneck for long-context inference and largely overlooked in on-device scenarios. It complements weight-centric optimizations to enable long-context LM inference on memory-limited devices.

%% file: sections/8.conclusion.tex
\section{Conclusion}

We have presented \SystemName, a KV cache offloading scheme to support long-context on-device LM inference on mobile and embedded devices. \SystemName leverages non-volatile secondary storage as a swapping medium to offload KV cache data. 
Through the co-design of algorithms and the runtime system, it delivers substantial system efficiency improvements over existing KV cache offloading solutions while maintaining generation quality. 
\SystemName delivers good scalability under different context lengths, batch sizes, model sizes, and limited memory budgets.

%% file: sections/9.appendix.tex
\appendix
\twocolumn
\nobalance
\setcounter{table}{0}
\setcounter{figure}{0}

\input{sections/exp_data/main_acc_qwen}

\input{sections/exp_data/main_throughput}

\begin{table}[t]
    \centering
    \captionsetup{skip=0pt}
    \footnotesize
    \caption{Accuracy (\%) w/ and w/o rolling buffer (RB) for different KV prediction/selection group sizes ($G$).}
    \label{tab:acc_rb}
    \begin{tabular}{l|cccc}
        \toprule
          & G=2 & G=4 & G=8 & G=12 \\
        \midrule
         With RB & 86.9 & 86.6 & 84.9 & 84.4\\
         No RB & 57.9 & 41.1 & 34.9 & 30.6 \\
        \bottomrule
    \end{tabular}
\end{table}

\section{Details of \DesignName's Parameter Search\label{details_para_search}}
As described in the main text, \SystemName provides an API to search for a set of parameters to be used by the \SystemName runtime. Search is achieved through a combination of heuristics and a greedy-based search parameter solver using profiling informaiton.

\subsection{Precomputed parameter lookup tables\label{sec:lookup}}
To reduce tuning cost, the \SystemName API allows precomputing hardware-independent parameters on a high-performance server. This is done by profiling the LM on selected datasets, with results stored in two \emph{lookup tables}: (1) reuse buffer capacity $\mathcal{C}$ to reuse rate, and (2) compression ratio $\sigma$ to the low-rank adapter for building the compressed K cache. As shown in Sec.\ref{sec:analysis}, reuse rates for a given $\mathcal{C}$ are largely input-invariant, so we store the average value to accelerate tuning. By default, precomputation uses the C4 dataset~\cite{raffel2020exploring} with 20 sampled batches, though both dataset and sample size are user-configurable via the API.

\subsection{Number of selected entries\label{sec:nse}} Recall that \SystemName predicts and loads important KV cache entries in groups. For a group size $G$ and a selected group number $M$, the size of the KV preloading buffer is $H_{kv}MG$, where $H_{\mathrm{kv}}$ is the number of KV heads. While $H_{kv}MG$ affects the I/O latency, its impact on memory is negligible because: (a) our layerwise preloading strategy allows this buffer to be shared across all layers, and with KV reuse enabled, the buffer is merged into the reuse buffer; and (b) prior work~\cite{zhang2023h2o,sun2024shadowkv} shows that critical KV entries are sparse, typically comprising less than 5\% of the long context. This allows us to preset $H_{kv}MG$ to reduce the search space with minimal accuracy loss. As $H_{kv}$ is a model constant, we control the number of selected entries by fixing $MG=\texttt{Const}$ (e.g., $\texttt{Const}=400$).

\subsection{Metric measurement\label{sec:mm}} We use sampled profiling to obtain two metrics, \emph{I/O delay} and \emph{model delay}, which depend on the underlying hardware and the target LM. In this work, we configure \SystemName to use NVIDIA NVTX~\cite{nvidia_nvtx} and Nsight Systems~\cite{nvidia_nsight_systems} to collect timing information for the core functions (instrumentation is enabled by turning on an \SystemName option) of the \SystemName runtime, but other profiling tools can be used by overwriting a \SystemName method. 
Profiling is performed by sampling different combinations, $(b, S)$, of batch sizes $b$, and context lengths $S$. 
Specifically, we construct the sets $\{b\}$ and $\{S\}$ by sweeping $b \in [1, b_{\max}]$ and $S \in [S_{\min}, S_{\max}]$, where $S_{\min}$ is a predefined lower bound context length (e.g., 4K). By default, the steps for $b$ and $S$ are set to $1$ and $2\mathrm{K}$, respectively, but may be increased when $b_{\max}$ or $S_{\max}$ is excessively large to reduce the time cost of exhaustive profiling. We apply interpolation on the measured configurations to estimate the values of missing points.   In addition to $(b, S)$, we sweep the compression ratio $\sigma$ and reuse buffer capacity $\mathcal{C}$ over the keys of lookup tables (Sec.~\ref{sec:lookup}), which are designed to span a sufficiently large range and fine granularity. 

Specifically, for given $b$, $S$, $\sigma$, $\mathcal{C}$, the number of selected groups $M$, group size $G$, and $MG=\texttt{Const}$, we profile \SystemName when serving the target LM on a sample data:

\cparagraph{The I/O delay:} $T_{\text{io}}(b, \texttt{Const}, G, \mathcal{C})$ measures disk load time for KV entries.
    The effect of $\mathcal{C}$ is modeled by averaging the latency using multiple randomly sampled reuse patterns according to the lookup reuse rate.
    We omit incremental disk updates because their latency is small and hidden within the compute–I/O pipeline. 

\cparagraph{The model delay:} $T_{\text{model}}(b, \texttt{Const}, \mathcal{C}, S, \sigma)$ covers the time spent on standard transformer blocks (i.e., attention and FFN), and the overhead introduced by \SystemName's prediction and reuse-buffer management. Attention time depends on $b$ and $\texttt{Const}$; prediction time on $b$, $\sigma$, and $S$; and reuse management cost on $b$, $\mathcal{C}$, and $\texttt{Const}$. Compact K cache and rolling buffer updates are negligible. 

As a Transformer-based LM typically consists of multiple stacked Transformer blocks, we only need to profile a single block as a representative, which takes only a few seconds per parameter combination.

\begin{figure}[t]
  \centering
  \captionsetup{skip=0pt}
  \includegraphics[width=\linewidth]{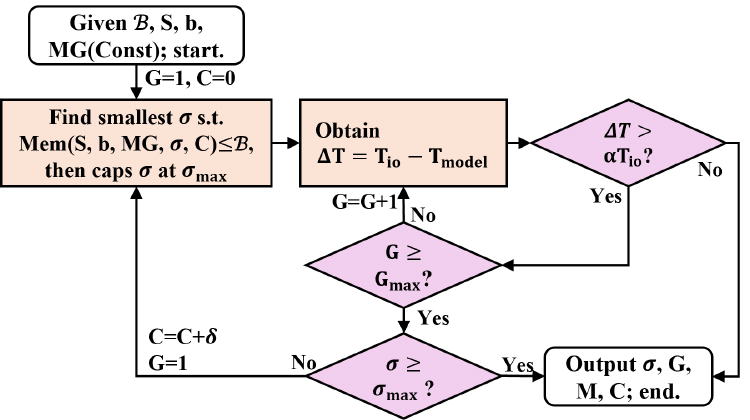}
  \caption{Workflow of the \SystemName runtime parameter solver for parameter tuning.}
  \label{fig:solver}
\end{figure}

\subsection{Parameter solver}

Figure~\ref{fig:solver} shows the workflow of the \SystemName parameter solver. It first determines the compression ratio $\sigma$ so that the memory budget $\mathcal{B}$ is not exceeded. It then finds the smallest possible $G$ that can hide I/O latency within computation.

\cparagraph{Overlapping I/O and computation.} I/O latency can be substantial on slower disks (e.g., eMMC) or under batched processing, and fully hiding I/O may severely degrade generation quality. To balance this trade-off, we introduce a relaxation factor $\alpha$, permitting only a $(1-\alpha)$ fraction of I/O to be hidden within computation. If hiding fails at $G_{\max}$, the solver reallocates part of the memory budget to the reuse buffer (increasing $\mathcal{C}$ by $\delta$) to reduce I/O volume, while adjusting $\sigma$ to stay within budget and restarting the search from $G=1$. The search stops once $(1-\alpha)$ of I/O is overlapped with computation or the limits $\sigma_{\max}$ and $G_{\max}$ are reached. Generation quality is preserved by keeping $\sigma$ and $G$ as small as possible, with $\sigma_{\max}$ and $G_{\max}$ preventing excessive accuracy degradation under tight memory budgets.

\cparagraph{Record solutions.} \SystemName records the optimal parameter setting for each batch size $b$ and context length $S$ pair $(b, S)$.Given $\mathcal{B}$ and $MG=\text{Const}$, the solver sweeps over $(b, S)$ pairs and stores every solution. At runtime, parameters are retrieved by exact match; if unavailable, the nearest pair is chosen. Since users only specify the maximum memory budget $\mathcal{B}{\max}$ and the maximum batch size $b{\max}$, actual memory usage scales with $b$. To account for this, we allocate a uniform per-batch budget of $\mathcal{B}{\max}/b{\max}$.

\section{More Evaluation Results\label{app:more_eval}}

Tab.~\ref{tab:main_acc_qwen} compares the generation accuracy of Qwen3-8B on RULER and LongBench.
Tab.~\ref{tab:full_throughput} presents the generation throughput on NVMe and eMMC with batch sizes of 1, 2, 4, 8, 16, and context lengths of 8K, 16K, 24K and 32K for LLaMA3-8B.
Tab.~\ref{tab:acc_rb} compares the average generation accuracy with and without the rolling buffer on KV entry group sizes ($G$) of 2, 4, 8, and 12, for LLaMA3-8B over MV, QA1, and VT datasets from RULER.

%% file: sections/exp_data/main_acc_qwen.tex
\begin{table*}
\captionsetup{skip=0pt}
\centering
\caption{Relative accuracy loss (\%) of Qwen3-8B compared to Full-KV on RULER (left) and LongBench (right).}
\label{tab:main_acc_qwen}
\footnotesize
\setlength{\tabcolsep}{1pt}

\begin{tabular}{l|ZZZZZZZZZ|ZZZZZZZ}
\toprule
\textbf{Methods}
& \textbf{S1} & \textbf{S2} & \textbf{MK1} & \textbf{MQ} & \textbf{MV}
& \textbf{QA1} & \textbf{QA2} & \textbf{VT} & \textbf{Avg.}
& \textbf{SQA} & \textbf{MQA} & \textbf{SUM} & \textbf{FSL} & \textbf{SYN} & \textbf{COD} & \textbf{Avg.} \\
\midrule


\cellcolor{lightblue}Full-KV (raw \%)& \cellcolor{lightblue}100.0 & \cellcolor{lightblue}100.0 & \cellcolor{lightblue}100.0 & \cellcolor{lightblue}99.75 & \cellcolor{lightblue}98.75 & \cellcolor{lightblue}79.0 & \cellcolor{lightblue}67.0 & \cellcolor{lightblue}100.0 & \cellcolor{lightblue}93.06 & \cellcolor{lightblue}41.63 & \cellcolor{lightblue}44.61 & \cellcolor{lightblue}26.02 & \cellcolor{lightblue}69.84 & \cellcolor{lightblue}100.0 & \cellcolor{lightblue}56.86 & \cellcolor{lightblue}56.49 \\
\cline{1-17}
InfiniGen   & -100.0 & -100.0 & -100.0 & -99.75 & -98.75 & -67.0 & -48.0 & -99.8 & -89.16 & -29.74 & -33.37 & -18.68 & -32.54 & -91.29 & -46.71 & -42.05 \\
InfiniGen*  & -56.0  & -100.0 & -99.0  & -99.75 & -98.75 & -34.0 & -18.0 & -38.2 & -67.96 & -11.13 & -6.64  & -4.66  & -8.31  & -41.0  & -13.64 & -14.23 \\
Loki        & \bfseries0.0   & \bfseries0.0   & \bfseries0.0   & -1.0   & -6.25  & -5.0  & \bfseries-5.0  & -3.0  & -2.53  & -2.48  & \bfseries{+0.5} & \bfseries{+1.2} & -0.86  & \bfseries0.0   & -3.74  & \bfseries-0.9 \\
ShadowKV    & -2.0  & -2.0  & -6.0  & -4.5   & -17.5  & -13.0 & -12.0 & -19.6 & -9.57  & -4.57  & -1.25  & -0.71  & -0.8 & -2.0   & -4.6   & -2.32 \\
\cellcolor{lightgreen}\DesignName\textsuperscript{NVMe} & \cellcolor{lightgreen}\bfseries0.0 & \cellcolor{lightgreen}\bfseries0.0 & \cellcolor{lightgreen}\bfseries0.0 & \cellcolor{lightgreen}-0.5 & \cellcolor{lightgreen}\bfseries{+0.5} & \cellcolor{lightgreen}\bfseries-2.0 & \cellcolor{lightgreen}\bfseries-5.0 & \cellcolor{lightgreen}\bfseries0.0 & \cellcolor{lightgreen}\bfseries-0.9 & \cellcolor{lightgreen}-1.3 & \cellcolor{lightgreen}-0.70 & \cellcolor{lightgreen}-1.05 & \cellcolor{lightgreen}-1.3 & \cellcolor{lightgreen}\bfseries0.0 & \cellcolor{lightgreen}\bfseries-1.2 & \cellcolor{lightgreen}\bfseries-0.9 \\
\cellcolor{lightgreen}\DesignName\textsuperscript{eMMC} & \cellcolor{lightgreen}\bfseries0.0 & \cellcolor{lightgreen}\bfseries0.0 & \cellcolor{lightgreen}\bfseries0.0 & \cellcolor{lightgreen}\bfseries-0.3 & \cellcolor{lightgreen}{+0.2} & \cellcolor{lightgreen}-4.0 & \cellcolor{lightgreen}-9.0 & \cellcolor{lightgreen}-0.4 & \cellcolor{lightgreen}-1.7 & \cellcolor{lightgreen}\bfseries-0.9 & \cellcolor{lightgreen}-1.25 & \cellcolor{lightgreen}-1.1 & \cellcolor{lightgreen}\bfseries-0.6 & \cellcolor{lightgreen}\bfseries0.0 & \cellcolor{lightgreen}-2.4  & \cellcolor{lightgreen}-1.1 \\
\cline{1-17}
Loki-t      & -100.0 & -100.0 & -100.0 & -99.75 & -98.75 & -59.0 & -44.0 & -93.8 & -86.91 & -14.07 & -10.87 & -4.84 & -8.29 & -25.5 & -27.12 & -15.11 \\
ShadowKV-t  & -100.0 & -98.0  & -96.0  & -97.0  & -96.25 & -38.0 & -24.0 & -100.0 & -81.15 & -15.68 & -15.26 & -8.38 & -10.33 & -44.0 & -16.0 & -18.27 \\
\cellcolor{lightgreen}\DesignName-t\textsuperscript{NVMe} & \cellcolor{lightgreen}\bfseries0.0 & \cellcolor{lightgreen}\bfseries0.0 & \cellcolor{lightgreen}\bfseries0.0 & \cellcolor{lightgreen}\bfseries0.0 & \cellcolor{lightgreen}-1.3 & \cellcolor{lightgreen}\bfseries-15.0 & \cellcolor{lightgreen}-12.0 & \cellcolor{lightgreen}\bfseries-0.2 & \cellcolor{lightgreen}\bfseries-3.6 & \cellcolor{lightgreen}\bfseries-1.8 & \cellcolor{lightgreen}-2.7 & \cellcolor{lightgreen}\bfseries-1.4 & \cellcolor{lightgreen}\bfseries-1.7 & \cellcolor{lightgreen}\bfseries0.0 & \cellcolor{lightgreen}-2.9 & \cellcolor{lightgreen}\bfseries-1.8 \\
\cellcolor{lightgreen}\DesignName-t\textsuperscript{eMMC} & \cellcolor{lightgreen}\bfseries0.0 & \cellcolor{lightgreen}\bfseries0.0 & \cellcolor{lightgreen}\bfseries0.0 & \cellcolor{lightgreen}-0.5 & \cellcolor{lightgreen}\bfseries-1.0 & \cellcolor{lightgreen}-21.0 & \cellcolor{lightgreen}\bfseries-11.0 & \cellcolor{lightgreen}-1.6 & \cellcolor{lightgreen}-4.4 & \cellcolor{lightgreen}-2.3 & \cellcolor{lightgreen}\bfseries-1.6 & \cellcolor{lightgreen}\bfseries-1.4 & \cellcolor{lightgreen}-2.6 & \cellcolor{lightgreen}\bfseries0.0 & \cellcolor{lightgreen}\bfseries-2.7 & \cellcolor{lightgreen}\bfseries-1.8 \\
\bottomrule
\end{tabular}
\end{table*}

%% file: sections/exp_data/main_throughput.tex
\begin{table*}[!htbp]
\captionsetup{skip=0pt}
\centering
\caption{Full throughput (tokens/s) comparison of LLaMA3-8B across batch sizes and context lengths (CLs).
}
\label{tab:full_throughput}
\footnotesize
\setlength{\tabcolsep}{2pt}
\begin{tabular}{ll|SSSSS|SSSSS|SSSSS|SSSSS}
\toprule
\multirow{2}{*}{\textbf{Disks}} & \multirow{2}{*}{\textbf{Methods}} 
& \multicolumn{5}{c}{\textbf{CL=8K}} 
& \multicolumn{5}{c}{\textbf{CL=16K}} 
& \multicolumn{5}{c}{\textbf{CL=24K}} 
& \multicolumn{5}{c}{\textbf{CL=32K}} \\
\cmidrule(lr){3-7} \cmidrule(lr){8-12} \cmidrule(lr){13-17} \cmidrule(lr){18-22}
& & \textbf{b=1} & \textbf{b=2} & \textbf{b=4} & \textbf{b=8} & \textbf{b=16}
  & \textbf{b=1} & \textbf{b=2} & \textbf{b=4} & \textbf{b=8} & \textbf{b=16}
  & \textbf{b=1} & \textbf{b=2} & \textbf{b=4} & \textbf{b=8} & \textbf{b=16}
  & \textbf{b=1} & \textbf{b=2} & \textbf{b=4} & \textbf{b=8} & \textbf{b=16} \\
\midrule

\multirow{7}{*}{eMMC} 
 & \cellcolor{gray!20}FlexGen &\cellcolor{gray!20}0.22 &\cellcolor{gray!20}0.22 &\cellcolor{gray!20}0.22 &\cellcolor{gray!20}0.22 &\cellcolor{gray!20}0.22 &\cellcolor{gray!20}0.11 &\cellcolor{gray!20}0.1 &\cellcolor{gray!20}0.11 &\cellcolor{gray!20}0.11 &\cellcolor{gray!20}0.11 &\cellcolor{gray!20} 0.08&\cellcolor{gray!20} 0.08 &\cellcolor{gray!20} 0.08 &\cellcolor{gray!20}0.08&\cellcolor{gray!20} 0.07 &\cellcolor{gray!20}0.05 &\cellcolor{gray!20}0.05 &\cellcolor{gray!20}0.05 &\cellcolor{gray!20}0.05 &\cellcolor{gray!20}0.05 \\
 & Loki/InfiniGen & 0.1 & 0.1 & 0.1 & 0.1 & 0.1 & 0.1 & 0.1 & 0.1 & 0.1 & 0.1 & 0.1 & 0.1 & 0.1 & 0.1& 0.12 & 0.1 & 0.1 & 0.1 & 0.1 & 0.1\\
 & \cellcolor{gray!20}InfiniGen* & \cellcolor{gray!20}1.74 & \cellcolor{gray!20}1.64 & \cellcolor{gray!20}1.46 & \cellcolor{gray!20}1.54 & \cellcolor{gray!20}1.51 & \cellcolor{gray!20}1.71 & \cellcolor{gray!20}1.65 & \cellcolor{gray!20}1.46 & \cellcolor{gray!20}1.44 &\cellcolor{gray!20}1.40 & \cellcolor{gray!20}1.63 & \cellcolor{gray!20}1.50 & \cellcolor{gray!20}1.44 & \cellcolor{gray!20}1.35 & \cellcolor{gray!20}1.35 & \cellcolor{gray!20}1.61 & \cellcolor{gray!20}1.51 & \cellcolor{gray!20}1.38 & \cellcolor{gray!20}1.35 & \cellcolor{gray!20}1.35\\
 & InfiniGen*+ru &4.30 & 5.22 &5.66 &4.58 &3.53 & 3.96 & 4.69 &5.24 &4.23 &3.71 &4.15 & 4.5 & 5.07 & 4.4 & 3.5 & 4.2 & 4.4 & 4.9 & 4.5 & 3.5\\
 & \cellcolor{gray!20}ShadowKV & \cellcolor{gray!20}2.73 & \cellcolor{gray!20}3.68 & \cellcolor{gray!20}4.21 & \cellcolor{gray!20}4.45 & \cellcolor{gray!20}4.42 & \cellcolor{gray!20}2.97 & \cellcolor{gray!20}3.76 & \cellcolor{gray!20}4.14 & \cellcolor{gray!20}4.40 & \cellcolor{gray!20}3.44 & \cellcolor{gray!20}2.83 & \cellcolor{gray!20}3.54 & \cellcolor{gray!20}4.11 & \cellcolor{gray!20}4.11 & \cellcolor{gray!20}3.37 & \cellcolor{gray!20}2.97 & \cellcolor{gray!20}3.72 & \cellcolor{gray!20}4.17 & \cellcolor{gray!20}3.76 & \cellcolor{gray!20}3.4 \\
 & \cellcolor{lightgreen}\DesignName & \cellcolor{lightgreen}\bfseries5.89 & \cellcolor{lightgreen}\bfseries10.76 & \cellcolor{lightgreen}\bfseries16.07 & \cellcolor{lightgreen}\bfseries18.67 &\cellcolor{lightgreen}\bfseries15.03 &\cellcolor{lightgreen}\bfseries5.94 &\cellcolor{lightgreen}\bfseries10.45 &\cellcolor{lightgreen}\bfseries16.16 &\cellcolor{lightgreen}\bfseries15.79 &\cellcolor{lightgreen}\bfseries11.18 &\cellcolor{lightgreen} \bfseries6.02 &\cellcolor{lightgreen} \bfseries10.35 & \cellcolor{lightgreen}\bfseries15.27 &\cellcolor{lightgreen} \bfseries15.68 &\cellcolor{lightgreen}\bfseries11.1 &\cellcolor{lightgreen} \bfseries5.96&\cellcolor{lightgreen}\bfseries10.25  &\cellcolor{lightgreen}\bfseries15.19 &\cellcolor{lightgreen}\bfseries15.69 &\cellcolor{lightgreen}\bfseries10.9 \\
\midrule

\multirow{7}{*}{NVMe} 
 & \cellcolor{gray!20}FlexGen &\cellcolor{gray!20}1.6 &\cellcolor{gray!20}1.6 &\cellcolor{gray!20}1.6 &\cellcolor{gray!20}1.6 &\cellcolor{gray!20}1.6 &\cellcolor{gray!20}0.8 &\cellcolor{gray!20}0.8 &\cellcolor{gray!20}0.8 &\cellcolor{gray!20}0.8 &\cellcolor{gray!20}0.8 &\cellcolor{gray!20} 0.50 &\cellcolor{gray!20} 0.50 & \cellcolor{gray!20}0.48 &\cellcolor{gray!20} 0.45 &\cellcolor{gray!20} 0.45 &\cellcolor{gray!20}0.4 &\cellcolor{gray!20}0.4 &\cellcolor{gray!20}0.4 &\cellcolor{gray!20}0.4 &\cellcolor{gray!20}0.4 \\
 & Loki/InfiniGen & 1.9 & 1.9 & 1.9 & 1.9 & 1.9 & 1.9 & 1.9 & 1.9 & 1.9 & 1.9 & 1.9 & 1.9 & 1.9 & 1.9 & 1.8 & 1.9 & 1.9 & 1.9 & 1.8&1.8 \\
 & \cellcolor{gray!20}InfiniGen* & \cellcolor{gray!20}5.18 & \cellcolor{gray!20}8.77 & \cellcolor{gray!20}10.86 & \cellcolor{gray!20}12.36 & \cellcolor{gray!20}13.91 & \cellcolor{gray!20}4.95 & \cellcolor{gray!20}8.36 & \cellcolor{gray!20}10.81 & \cellcolor{gray!20}11.32 & \cellcolor{gray!20}13.1 & \cellcolor{gray!20}5.3 & \cellcolor{gray!20}8.15 & \cellcolor{gray!20}11.00 & \cellcolor{gray!20}11.78 & \cellcolor{gray!20}13.5 &\cellcolor{gray!20}5.1 &\cellcolor{gray!20}7.52 &\cellcolor{gray!20}8.88 &\cellcolor{gray!20}10.71 &\cellcolor{gray!20}11.96 \\
 & InfiniGen*+ru & 5.46 & 9.64 & 16.75 & 24.97 & 30.77 & 5.22 & 9.15 & 14.86 & 22.50 &26.57 & 5.15 & 8.72 & 13.05 &19.42 & 23.1 & 4.93 & 8.5 & 11.89 & 14.33 & 17.22 \\
 &\cellcolor{gray!20}ShadowKV & \cellcolor{gray!20}6.47 &\cellcolor{gray!20} 11.14 &\cellcolor{gray!20} 16.9 & \cellcolor{gray!20}22.2 & \cellcolor{gray!20}26.69 & \cellcolor{gray!20}6.41 &\cellcolor{gray!20} 10.43 &\cellcolor{gray!20} 16.31 &\cellcolor{gray!20} 21.85 &\cellcolor{gray!20} 26.71 &\cellcolor{gray!20} 6.38 &\cellcolor{gray!20} 9.95 &\cellcolor{gray!20} 16.05 &\cellcolor{gray!20} 20.41 &\cellcolor{gray!20} 25.81 &\cellcolor{gray!20} 6.38 &\cellcolor{gray!20} 10.03 &\cellcolor{gray!20} 16.25 &\cellcolor{gray!20} 21.49 &\cellcolor{gray!20} 26.21\\
 
 & \cellcolor{lightgreen}\DesignName & \cellcolor{lightgreen}\bfseries6.78 & \cellcolor{lightgreen}\bfseries11.95 &\cellcolor{lightgreen} \bfseries19.98 &\cellcolor{lightgreen} \bfseries35.74 &\cellcolor{lightgreen}\bfseries48.16 &\cellcolor{lightgreen}\bfseries6.93 &\cellcolor{lightgreen}\bfseries11.91 &\cellcolor{lightgreen}\bfseries20.80 &\cellcolor{lightgreen}\bfseries35.05 &\cellcolor{lightgreen}\bfseries46.06 &\cellcolor{lightgreen}\bfseries7.00 &\cellcolor{lightgreen}\bfseries11.79 &\cellcolor{lightgreen}\bfseries21.34 &\cellcolor{lightgreen}\bfseries35.84 &\cellcolor{lightgreen}\bfseries45.15 &\cellcolor{lightgreen}\bfseries6.85 &\cellcolor{lightgreen}\bfseries11.9 &\cellcolor{lightgreen}\bfseries21.41 &\cellcolor{lightgreen}\bfseries35.6 &\cellcolor{lightgreen}\bfseries46.79 \\
\midrule

None & \cellcolor{lightblue}vLLM &\cellcolor{lightblue} 10.03 & \cellcolor{lightblue}19.28 & \cellcolor{lightblue}35.43 &\cellcolor{lightblue} 55.23 &\cellcolor{lightblue} 81.71 &\cellcolor{lightblue} 9.69 &\cellcolor{lightblue} 17.14 &\cellcolor{lightblue} 28.38 &\cellcolor{lightblue} 41.19 &\cellcolor{lightblue} 39.46 &\cellcolor{lightblue} 9.19 &\cellcolor{lightblue} 14.80 &\cellcolor{lightblue} 23.96 &\cellcolor{lightblue} 32.06 &\cellcolor{lightblue} 30.02 &\cellcolor{lightblue} 9.22 &\cellcolor{lightblue} 14.06 &\cellcolor{lightblue} 21.03 &\cellcolor{lightblue} 20.77 &\cellcolor{lightblue} 23.21 \\
\bottomrule
\end{tabular}

\end{table*}